\documentclass[12pt]{article}

\usepackage{amsmath,amsfonts}
\usepackage{amssymb}
\usepackage{hhline}
\usepackage{epsfig,cite}
\usepackage{stmaryrd}
\usepackage[usenames,dvips]{color}
\usepackage{fullpage}
\usepackage{verbatim}
   
\makeatletter
\@addtoreset{equation}{section}
\makeatother

\newcommand{\be}{\begin{equation}}
\newcommand{\ee}{\end{equation}}
\newcommand{\bea}{\begin{eqnarray}}
\newcommand{\eea}{\end{eqnarray}}

\newcommand{\tr}{\operatorname{tr}}

\newcommand{\h}{\mathfrak h}

\newcommand{\nn}{\nonumber}

\newcommand{\M}{\mathcal M}
\newcommand{\OO}{\mathcal O}

\newcommand{\wz}{{m}_{11}^2}
\newcommand{\hz}{m_{12}^2}
\newcommand{\oz}{m_{22}^2}

\begin{document}

\thispagestyle{empty}

\begin{center}
\hfill BI-TP 2013/05\\
\hfill UAB-FT-730
\begin{center}

\vspace{.5cm}

{\Large\sc A Light Supersymmetric Higgs Sector Hidden by \\ \vspace{0.3cm} 

 a Standard Model-like Higgs}

\end{center}

\vspace{1.cm}

\textbf{ Antonio Delgado$^{\,a}$, Germano Nardini$^{\,b}$,
and Mariano Quiros$^{\,c}$}\\

\vspace{1.cm}
${}^a\!\!$ {\em {Department of Physics, University of Notre Dame,\\Notre Dame, IN 46556, USA}}

\vspace{.1cm}
${}^b\!\!$ {\em {Fakult\"at f\"ur Physik, Universit\"at Bielefeld,
    D-33615 Bielefeld, Germany}}

\vspace{.1cm}

${}^c\!\!$ {\em {Instituci\'o Catalana de Recerca i Estudis  
Avan\c{c}ats (ICREA) and\\ Institut de F\'isica d'Altes Energies, Universitat Aut{\`o}noma de Barcelona\\
08193 Bellaterra, Barcelona, Spain}}

\end{center}

\vspace{0.8cm}

\centerline{\bf Abstract}
\vspace{2 mm}
\begin{quote}\small
  Extending the Higgs sector of the MSSM by triplets alleviates the
  little hierarchy problem and naturally allows for enhancements in
  the diphoton decay rate of the lightest CP-even Higgs $h$.  In the
  present paper we analyze in detail the Higgs phenomenology of this
  theory with $m_h\simeq 126\,$GeV. We mostly focus on a light Higgs
  sector where the pseudoscalar $A$, the next-to-lightest CP-even
  scalar $H$ and the charged $H^\pm$ Higgses are naturally at the
  electroweak scale. It turns out that for any value $m_A\gtrsim m_h$
  there is a parameter region at small $\tan\beta$ where the CP-even
Higgs sector appears at colliders as the SM one, except for loop-induced corrections. Notably the existence of this SM-like point, which is absent in the MSSM, is
  shared with supersymmetric theories where there are extra
  singlets. We also highlight a second parameter region at small $m_A$
  and small $\tan\beta$ where the $h$ signal strengths, diphoton
  channel included, are SM-like except those of bottoms and taus which
  can have at most a 10-15\% splitting. Improvements in the $A$ and
  $H^\pm$ searches are worthwhile in order to discriminate this
  scenario from the SM.
 \end{quote}

\vfill

 \newpage
 \section{Introduction}
 \label{introduction}
 
 The ATLAS and CMS collaborations at CERN are currently finding no
 clear discrepancies between the collected data and the predictions of
 the Standard Model (SM) with a Higgs mass around
 126\,GeV~\cite{Atlas,CMS}. Any firm conclusion about the lack of
 relevant new physics in the LHC data is however impossible because of
 the experimental uncertainties still being too large at present. Many
 theoretical works have thus elaborated plausible SM extensions in
 view that some of the present experimental anomalies will be
 confirmed with less statistical error, as for instance the
 $\gamma\gamma$, $bb$ and $\tau\tau$ Higgs rates. In this sense many
 efforts have been made to extend the
 SM~\cite{Akeroyd:2012ms,Carena:2012xa,Wang:2012ts,Chun:2012jw,Chala:2012af}
 or to look for appropriate parameter regions in the minimal (MSSM) or
 non-minimal supersymmetric SM extensions
 ~\cite{Carena:2012gp,Kitahara:2012pb,Basso:2012tr,Huo:2012tw,SchmidtHoberg:2012yy},
 as those including extra singlets
 (NMSSM)~\cite{Ellwanger:2011aa,Cao:2012fz,Vasquez:2012hn,Benbrik:2012rm,Gogoladze:2012jp,Choi:2012he}
 and triplets~\cite{Basak:2012bd,Delgado:2012sm,Kang:2013wm} (or even
 in the presence of effective operators beyond the
 MSSM~\cite{Boudjema:2012in}), whose original motivation, besides
 adjusting the LHC data, is to solve the hierarchy problem.
 
 Although the MSSM solves the hierarchy problem, i.e.~it provides a
 technical solution to the existence of a grand hierarchy between the
 GUT (or Planck) and the electroweak scales, it requires an
 (unpleasant) amount of fine tuning in the electroweak sector in order
 to reproduce the 125-126\,GeV Higgs mass, the so-called little
 hierarchy problem.  Non-minimal supersymmetric scenarios generically
 alleviate the little hierarchy problem since they provide an extra
 tree-level contribution to the SM-like Higgs mass which makes the
 theory less dependent on large radiative corrections and, in turn, on
 large values of the supersymmetry breaking scale. If one keeps the SM
 group $SU(3)_c\otimes SU(2)_L\otimes U(1)_Y$ as the gauge symmetry of
 the theory the only remaining possibility (assuming renormalizable
 couplings) is to introduce an extra sector of chiral superfields
 coupled to the MSSM Higgs sector in the superpotential. These
 extended models are then limited to involving only extra singlets
 and/or triplets with hypercharges $Y=0,\pm
 1$~\cite{Espinosa:1991gr,Espinosa:1991wt}. In comparison with singlet
 extensions, models with triplets present the extra bonus of having
 new charginos which can be strongly (but perturbatively) coupled to
 the Higgs sector and may radiatively rise the $\gamma\gamma$
 production rate above the SM prediction. Therefore, if these
 radiative corrections are small, as for instance when the new
 charginos are heavy enough, the phenomenology of the model is
 expected to be similar to that of supersymmetric theories with extra
 singlets weakly mixed with Higgs doublets.
 
 The simplest of such a MSSM extension, i.e.~a $Y=0$ triplet coupled
 to the Higgs sector with a superpotential coupling $\lambda$, was
 investigated in Ref.~\cite{Delgado:2012sm} in the decoupling limit,
 i.e.~for the CP-odd Higgs mass $m_A$ much larger than the electroweak
 scale. In this case the tree-level couplings of the lightest
 Higgs $h$ to the SM particles equal those of the SM and the major
 potential signature of the model is an enhancement in processes
 mediated by charginos, in particular the decay rate of the
 $h\to\gamma\gamma$ channel~\cite{Delgado:2012sm}. The decoupling
 regime however introduces a sizeable amount of fine-tuning in the
 model as, in the minimization conditions of the Higgs potential,
 cancellations of $\mathcal O (m_A^2)$ terms are required to fix the
 $Z$ boson squared mass $m_Z^2$ to its experimental value. It is
 therefore compelling to abandon the decoupling regime and consider
 $m_A$ at the electroweak scale.
 
 In this paper we make a detailed study of the $Y=0$ triplet extension
 of the MSSM, for a value of the light Higgs mass $m_h\simeq126$ GeV
 in the non-decoupling regime, where the masses of the remaining MSSM
 Higgs scalars $(H,A,H^\pm)$ are at the electroweak scale.  Stops
   and scalar triplets are instead assumed at the TeV scale in
   agreement with collider and electroweak precision
   observables~\cite{pdg}, and so their radiative corrections to the
   Higgs production and decay rates are consequently
   small~\cite{Delgado:2012sm, DiChiara:2008rg}. The most striking
 result of the paper is that for any value of $m_A\gtrsim m_h$
 there is a SM-like point at $\tan\beta=\tan\beta_c$ and
 $\lambda=\lambda_c$ (whose explicit value depends on the rest of
 supersymmetric parameters) where the tree-level Higgs couplings
 $g_{hXX}$ (with $X=W,\, Z,\, b,\, t,\, \tau,$ where $b,\, t,\, \tau$
 stand respectively for down quarks, up quarks and charged leptons of
 the three families) equal the SM values and only a departure in the
 branching ratio of $h\to\gamma\gamma$, as large as 40\%, can appear
 depending on the chargino spectrum. This means that the region around
 the SM-like point is consistent with actual experimental data. Given
 the present large experimental uncertainties no attempt has been made
 to fit the ATLAS and CMS measurements, a task which will be
 compelling in the future when experimental data will be more
 precise. Moreover, depending on the value of $m_A$, there can exist a
 second point at $\lambda>\lambda_c$ where the couplings $g_{hXX}$
 $(X=t,\, W,\, Z,\,\gamma)$ are similar to the SM ones, while there is
 a splitting between the $bb$ and $\tau\tau$ channels produced by
 radiative corrections. The region around this point will also deserve
 a more detailed analysis when more accurate experimental data will be
 available.

 The outlook of the paper is as follows. In section~\ref{model} we
 introduce the model. Its scalar sector (including the minimization
 conditions for the electroweak minimum and scalar masses including
 radiative corrections) and the chargino sector, which will be
 relevant for the diphoton production. In particular we pay particular
 attention to the appearance of the SM-like point already
 mentioned. In section~\ref{sec:coupl} the Higgs coupling are
 presented and it is explicitly shown how the branching ratios of the
 $bb,\, \tau\tau,\, tt,\, WW,$ and $ZZ$ channels normalized to their
 SM values are precisely equal to one at the SM-like point for any
 value of $m_A\gtrsim m_h$ while the actual value for the
 $\gamma\gamma$ channel presents an enhancement with respect to the SM
 value depending on the spectrum of chargino masses. The Higgs
 production rates are analyzed in detail in section~\ref{sec:4} where
 a particular example with $m_A=140$ GeV is exhibited as a function of
 $\lambda$ for both gluon-fusion and vector boson fusion production
 mechanisms. The appearance of the SM-like point, as well as the
 existence of the second point compatible with present experimental
 data, is made explicit. We also present results on the production of
 the next-to-lightest CP-even Higgs (with a mass around 140 GeV) that
 explain why this particle has been undetected in the present LHC
 data. All these results are obtained by means of an approximation
 dealing with the scalar triplet being decoupled from the Higgs
 doublets (fully justified if the scalar triplet is sufficiently
 heavy), which implies a tiny contribution to the $T$ parameter
 triggered by the triplet vacuum expectation value (VEV), and in
 section~\ref{sec:5} we check the accuracy of this approximation.
We find that in our analysis the approximation leads to a theoretical error on
the Higgs mass of $\pm 0.5$\,GeV, which is smaller than the expected
uncertainty from unconsidered radiative contributions. On the other hand the
error made in the determination of the Higgs couplings is less than
1\%. Finally in section~\ref{sec:6} we consider the full case where the
triplet can be rather light and get mixed with the doublet scalars as much as
the present experimental bounds on the $T$ parameter allows for. Instead of a
general analysis, which is outside the scope of the present paper, we focus on
the trajectory of the SM-like points corresponding to a scalar triplet mass
decreasing from large to smaller values in agreement with electroweak constraints. It turns out that the maximum allowed value of the
$\gamma\gamma$ production rate increases as the triplet scalar mass decreases
and it can reach at most a 40\% enhancement with respect to the SM
value. Finally some technical details concerning the existence of the SM-like
point are shown in the appendix.

 \section{The model}
 \label{model}
  A $Y=0$ triplet $\Sigma$ is described by its electrically charged and
 neutral components, $\xi_1^-,\xi_2^+$ and $\xi^0$, as
\be
\Sigma=\left(
\begin{array}{cc}
  \xi^0/\sqrt{2} & -\xi_2^+\\
  \xi_1^-&  -\xi^0/\sqrt{2}
\end{array}
\right)~.
\label{Sigma}
\ee
The most general renormalizable superpotential that couples $\Sigma$
to the Higgs sector is given by
\begin{equation}
\Delta W=\lambda H_1\cdot \Sigma H_2+\frac{1}{2}\mu_\Sigma \tr \Sigma^2+\mu H_1\cdot H_2~.\label{superpotential}
\end{equation}
where $A \cdot B \equiv \epsilon_{ij} A^iB^j$ with
$\epsilon_{21}=-\epsilon_{12}=1$ and $\epsilon_{22}=\epsilon_{11}=0$.
In the superpotential the identity Tr$(\Sigma^3)\equiv0$ prevents the
presence of the cubic term involving $\xi^{0}$ and $\xi^{\pm}$.
The new parameters (including soft terms) with respect to the MSSM are then, the
superpotential coupling $\lambda$, the supersymmetric mass
$\mu_\Sigma$, the soft-breaking masses $m_4$ and $B_\Sigma$ and the trilinear
soft-breaking parameter $A_\lambda$, and we assume no $CP$ violation in the Higgs sector.

\subsection{The scalar triplet-Higgs sector}
\label{subsec:sector}
The tree-level potential for the neutral components of the Higgs and
triplet sector, $H_{1,2}^0$ and $\xi^0$, is given by~\footnote{The
  analysis of the full triplet-Higgs potential including the
  components $\xi_1^-,\xi_2^+,H_1^-$ and $H_2^+$ is quite cumbersome.
  It can be easily imposed however that the physical minimum where
  none of these electrically-charged fields acquires a VEV is stable
  or at least long-lived~\cite{Espinosa:1991wt}. We can
  then carry out perturbation theory around this minimum.}
 \bea
 V&=&m_1^2 |H_1^0|^2+m_2^2 |H_2^0|^2+m_4^2 |\xi^0|^2 \nn\\
&+&\left|\mu H_2^0- \lambda H_2^0 \xi^0/\sqrt{2}\right|^2+\left|\mu
  H_1^0- \lambda H_1^0 \xi^0/\sqrt{2}\right|^2 \nn\\
&+&\left|\mu_\Sigma\xi^0-\lambda H_1^0 H_2^0/\sqrt{2}\right|^2+\frac{g^2+g'^2}{8}\left(|H_2^0|^2-|H_1^0|^2\right)^2 \nn\\
&+&\left(B_\Sigma\mu_\Sigma \xi^0\xi^0 -A_\lambda \lambda H_1^0 H_2^0\xi^0/\sqrt{2}-m_3^2 H_1^0H_2^0
  + {\rm h.c.}\right)~.
\label{potencial}
\eea
The experimental bound on the $T$-parameter constrains the parameters
of the potential. It requires the VEV of $\xi^0$ to be around or below
the GeV scale. Unless there is a fine tuning, this imposes the hierarchy
(cf.~e.g.~Ref.~\cite{Delgado:2012sm})
\be \displaystyle |A_\lambda|,\,|\mu| \,,|\mu_\Sigma| \lesssim
10^{-2}
\frac{m^2_\Sigma+\lambda^2 v^2/2}{\lambda v}~,\label{approximation}
\ee
with $m_\Sigma^2\equiv m_4^2+\mu_\Sigma^2+B_\Sigma\mu_\Sigma$.  Such a
hierarchy may naturally arise in some models of supersymmetry
breaking. In particular in gauge mediation~\cite{gaugemed} the
trilinear soft terms are loop suppressed with respect to soft breaking
masses. Moreover if the scale of supersymmetry breaking is high enough
and gauge interactions transmitting supersymmetry breaking are not the
SM ones, gravitational interactions can solve the $\mu$-problem while
the long running can create hierarchies among different soft
masses~\cite{antonio}. We will just assume this hierarchy in the rest
of the paper and consequently the small $\xi^0$ VEV will be neglected.

Unlike in the MSSM, the $D$-flat direction $|H_1^0|=|H_2^0|$ (with
$\xi^0=0$) is stable, independently of the values of the mass
parameters, due to the term $(\lambda^2/2) |H_1^0 H_2^0|^2$ in the
potential. This term also modifies the minimization conditions valid
for the MSSM. In the vacuum, where $\langle H_1^0\rangle=v_1$ and
$\langle H_2^0\rangle=v_2$, the (tree-level)
potential~\eqref{potencial} provides the correct electroweak symmetry
breaking if the following equalities are fulfilled:
\begin{eqnarray}
m_3^2&=&m_A^2 \sin\beta\cos\beta ~,\\
m_Z^2&=&\frac{m_2^2-m_1^2}{\cos 2\beta}-m_A^2+\frac{\lambda^2}{2}v^2~,\label{min_MZ}\\
m_A^2&=&m_1^2+m_2^2+2|\mu|^2+\frac{\lambda^2}{2}v^2~,\label{min_MA}
\end{eqnarray}
where $\tan\beta=v_2/v_1$,
$v=\sqrt{v_1^2+v_2^2}=174$ GeV, and $m_Z$ is
the $Z$ boson mass.

By the definitions $H_i^0=v_i+(h_i+i\chi_i)/\sqrt{2}$ and
$x=\,$Re$\,\xi^0/\sqrt{2}$, the CP-even scalar squared mass matrix can
be written as
\begin{eqnarray}
\frac{1}{2}(h_2,h_1,x)\, 
\widehat{\mathcal M}^2
\left( \begin{array}{c} h_2 \\ h_1 \\x\end{array}\right)
\qquad \textrm{with}\qquad
\widehat{\mathcal M}^2
=
\left( \begin{array}{ccc}
\text{\fontsize{21}{1em}\selectfont  $~\,{}_{\M^{{}^{{}_2}}}$} &  \begin{array}{c} \cdot \\ \cdot \end{array}\\
\begin{array}{cc} \cdot & \cdot \end{array} & {\,~m_\Sigma^2+\frac{\lambda^2}{2}v^2}
\end{array} \right)  
\label{matrix}~,
\end{eqnarray}
where $\M^2$ is, at tree level, given by
\be
\mathcal M^2_0=\left(
\begin{array}{cc}
m_A^2 \cos^2\beta+m_Z^2\sin^2\beta &(\lambda^2
v^2-m_A^2-m_Z^2)\sin\beta\cos\beta \\
(\lambda^2 v^2-m_A^2-m_Z^2)\sin\beta\cos\beta&m_A^2  
\sin^2\beta+m_Z^2\cos^2\beta 
\end{array}
\right)~.
\label{scalarmass}
\ee
The entries represented by dots are terms typically of
$\OO(\lambda\hat\mu v)$, with $\hat\mu= {\rm max}\{|\mu|,$
$|\mu_\Sigma|,|A_\lambda|\}$. If they are negligible with respect to
$m_\Sigma^2+\lambda^2v^2/2$, the diagonalization of
  $\widehat{\mathcal M}^2$ is practically independent of them and no
  mixing is present between the CP-even Higgs sector of the MSSM and
  the triplet. Moreover, under such a hierarchy, a similar splitting
  between the triplet and the CP-odd and charged MSSM Higgs sectors 
  also holds~\footnote{For explicit expressions of the corresponding mass
  matrices see e.g.~Ref.~\cite{Espinosa:1991wt,DiChiara:2008rg}.}. It
  then turns out that the phenomenology of the model can be described
  by quite simple analytic expressions since:
{\it i)} $\beta$ is the rotation angle diagonalizing the CP-odd and
charged Higgs squared-mass matrices, like in the MSSM; {\it ii)} The
parameter $m_A^2$ defined in eq.~\eqref{min_MA} is the tree-level
squared mass of the lightest CP-odd Higgs, and; {\it iii)} The
tree-level mass of the  lightest charged Higgs $H^\pm$ can be easily
expressed as~\footnote{Note that the $\lambda^2$-term in
  eq.~(\ref{chargedHiggs}) provides a positive contribution to
  $m_{H^\pm}^2$, unlike in singlet extensions of the MSSM. Therefore
  experimental lower limits on
  $m_{H^\pm}$~\cite{Heister:2002ev,Chatrchyan:2012vca,Aad:2012lea,Aad:2013hla}
  do not put any upper bound on $\lambda$ as a function of $m_A$.}
\be
m_{H^\pm}^2=m_A^2+m_W^2+\frac{\lambda^2}{2}v^2~.
\label{chargedHiggs}
\ee 

In order to understand analytically the main features of the
considered scenario we will focus on regimes where the entries
$\OO(\lambda\hat\mu v)$ in eq.~\eqref{matrix} can be ignored. Notice
that neglecting these off-diagonal terms leads to overestimating the
lightest eigenvalue of the squared mass matrix in eq.~\eqref{matrix},
$m_h^2$, by a correction of order of $\lambda^2 \hat\mu^2
v^2/(m_\Sigma^2m_h^2)$. {\it For the sake of comprehension} the
present analysis will thus focus on values of $m_\Sigma^2$ that are
sufficiently large to {\it safely} ensure this correction to be
negligible. Specifically for $\mu=\mu_\Sigma=250\,$GeV and $A_\lambda=
0$, the overestimate on $m_h=126\,$GeV is below 1 GeV when
$m_\Sigma\gtrsim 5\,$TeV~\footnote{These values of $\mu$ and
  $\mu_\Sigma$ are chosen in view of the Higgs diphoton rate
  enhancement and chargino bounds studied in
  section~\ref{sec:coupl}. The 1\,GeV estimate is instead obtained in
  the analysis of section~\ref{sec:5} including radiative
  corrections.}.  This parameter setting is widely consistent with the
bound~\eqref{approximation}, which would already be fulfilled at
$m_\Sigma\gtrsim 1.5\,$TeV. For the sake of analytic tractability,
however, we take hereafter $m_\Sigma= 5\,$TeV (with
$\mu=\mu_\Sigma=250\,$GeV and $A_\lambda= 0$).

Of course the matrix $\M^2$ differs from $\M_0^2$ by radiative
corrections, which also modify the tree-level minimization
conditions~\eqref{min_MZ} and~\eqref{min_MA}. We will consider
radiative corrections coming from fields strongly coupled to the Higgs
sector (but still in the perturbative regime) and with large
supersymmetry breaking masses, providing radiative corrections
enhanced by large enough logarithms (but still consistent with the
validity of the one-loop approximation). The most relevant
contributions are generated by the stop~\footnote{Corrections due to
  the sbottom sector are subleading in the phenomenologically
  interesting cases $\tan\beta\ll m_t/m_b$, where $m_t$ and $m_b$ are
  the top and bottom quark masses, respectively.} and
triplet~\footnote{Corrections coming from the Higgs sector are
  suppressed, with respect to those of the triplet, by small
  logarithms and they are then neglected throughout this paper.} sectors.
Their radiative corrections can be computed by means of the one-loop
effective potential in the presence of the $H_{1,2}^0$ background
fields.  Neglecting the Higgs-triplet mixing and assuming
$m_\Sigma^2\gg B_\Sigma\mu_\Sigma$, the background fields generate the
following scalar triplet and stop spectrum (at order $\lambda^2$ and
$h_t^2$):
\begin{eqnarray}
  m^2_{\xi^0}&=&m_\Sigma^2+\frac{\lambda^2}{2}\left(\left|
      H_1^0\right|^2+\left| H_2^0\right|^2\right)\nonumber~,\\
  m^2_{\xi^+_2}&=&m_\Sigma^2+\lambda^2\left| H_2^0\right|^2\nonumber~,\\
  m^2_{\xi^-_1}&=&m_\Sigma^2+\lambda^2\left| H_1^0\right|^2~,\nn\\
  m^2_{\tilde t}\,&=&m_Q^2+h_t^2\left| H_2^0\right|^2~,
\label{masasSigma}
\end{eqnarray}
where $h_t=m_t/(v \sin\beta)$ and the parameters $\mu$ and $A_t$ are
assumed to be much smaller than the (common) stop supersymmetry breaking
mass $m_Q$. 

By expanding the one-loop effective potential with the
above stop and triplet spectrum, it turns out that the tree-level
minimization conditions~\eqref{min_MA} have to be corrected by the
amount
\be
\Delta m_A^2 =\frac{3\lambda^2}{16\pi^2}m_\Sigma^2 (t_\Sigma-1)+\frac{3h_t^2}{8\pi^2}m_Q^2(t_Q-1)~,
\label{DeltamA}
\ee
with $t_\Sigma\simeq\log(m_\Sigma^2/\mu_\Sigma^2)$ and
$t_Q=\log(m_Q^2/m_t^2)$.  In the same way, $m_Z^2$ is given by
expression~\eqref{min_MZ}, where $m_A^2$ includes the radiative correction~\eqref{DeltamA}, plus the genuine radiative contribution
\be
\Delta m_Z^2 =-\frac{\lambda^4}{16\pi^2}v^2t_\Sigma+\frac{6h_t^2}{16\pi^2}\, \frac{m_Q^2}{\cos 2\beta}(t_Q-1)
\label{DeltamZ}~.
\ee
Moreover from the triplet sector the matrix $\M_0^2$ receives the
radiative contributions
\begin{eqnarray}
\Delta_\Sigma \mathcal M_{11}^2&=&\frac{5\lambda^4}{32\pi^2}t_\Sigma v^2\sin^2\beta\nonumber~,\\
\Delta_\Sigma \mathcal M_{22}^2&=&\frac{5\lambda^4}{32\pi^2}t_\Sigma v^2\cos^2\beta\nonumber~,
\\
\Delta_\Sigma \mathcal M_{12}^2&=&\frac{\lambda^4}{32\pi^2}t_\Sigma v^2\sin\beta\cos\beta~,
\label{radSigma}
\end{eqnarray}
while the stop sector provides the corrections~\footnote{In
  $\Delta_{\tilde t}\mathcal M_{11}^2$ and $\Delta_{\tilde t}\mathcal
  M_{12}^2$ some subleading terms are included for completeness
  (cf.~e.g.~\cite{Carena:1995bx}).  We instead omit $\Delta_{\tilde
    t}\mathcal M_{22}^2$ since it is negligible when
  $|\mu|^2,|A_t|^2\ll m_Q^2$ as bottom Yukawa, hypercharge and
    weak coupling contributions are subdominant to radiative
    corrections proportional to $\lambda$ and $h_t$ (both close to one
    in the parameter space we will be interested in).}
\begin{eqnarray}
\Delta_{\tilde t}\mathcal M_{11}^2&=&\frac{3}{8\pi^2}t_{Q}
h_t^2\sin^2\beta\left[-m_Z^2+2 h_t^2
  v^2\left(1+\frac{t_{Q}}{16\pi^2}\left(\frac{3h_t^2}{2}-8g_3^2\right)\right)\right]~,\nn\\
\Delta_{\tilde t}\mathcal M_{12}^2&=&\Delta_{\tilde t}\mathcal
M_{21}^2=\frac{3}{16\pi^2}h_t^ 2 \sin\beta\cos\beta\, m_Z^2\,
t_Q~.
\label{radstop}
\end{eqnarray}

The amount of fine tuning present in this scenario is very sensitive
to the choice of $m_\Sigma^2$, specially through the radiative
corrections to $m_A^2$. Indeed, the percent fine-tuning of $m_A^2$
with respect to $m_\Sigma^2$, which is defined as $100 (\partial \log
m_\Sigma^2/\partial\log m_A^2)$, does ameliorate from around 2\% to
40\% by lowering $m_\Sigma$ from 5 to 1.5\,TeV (for
$m_A=\mu_\Sigma=250\,$GeV). However, since our main conclusions,
obtained for $m_\Sigma=5\,$TeV, also cover more natural choices of
$m_\Sigma$ (see section~\ref{sec:6}) we perform the analysis at
$m_\Sigma=5$\,TeV keeping in mind that the fine tuning of the scenario
could be easily reduced by decreasing $m_\Sigma$~\footnote{We cannot
  consider $m_\Sigma\gg 5\,$TeV in the one-loop approximation. For
  that case we should improve the triplet radiative corrections in
  order to avoid perturbative problems.}.

On the other hand, $m_Z^2$ is little sensitive to $m_\Sigma^2$ once
$m_A$ has been set ($\Delta m_Z^2$ is just logarithmically dependent
on $m_\Sigma$).  In order to fix $m_Z^2$ to its experimental value
only a tuning due to $m_Q^2$ is then required, and naturalness
criterion consequently drives $m_Q$ to its lower bound coming from
direct searches of stops [unlike in the MSSM, here the constraint
$m_h\simeq 126\,$GeV does not impose a stringent bound on $m_Q^2$
(even for $A_t=0$) because of the tree-level interaction
$\lambda^2|H_1|^2|H_2|^2$ ]. As a rather natural value, in the rest of
the paper we take $m_Q=m_U=700\,$ GeV, which is in full agreement with
experimental constraints if gluinos are not
light~\cite{Lungwitz:2012sza}.

In summary, and unless explicitly specified, the parameter setting
considered in this analysis is
\bea
&&m_Q=m_U=700\,{\rm GeV}~, \quad A_t=0~, \quad m_\Sigma=5\,{\rm TeV}~,\nn \\
&&\mu=\mu_\Sigma=250\,{\rm GeV}~,\quad ~~|M_3|=1\,{\rm TeV}~.
\label{setting}
\eea

\subsection{The constraint $\mathbf{m_h\simeq 126}$\,GeV and the
  SM-like point}
  \label{subsec:constraint}

  In view of the recent ATLAS and CMS results~\cite{Atlas,CMS} we
  impose in the model the constraint that the lightest Higgs boson
  mass is around 126\,GeV~\footnote{Several studies have investigated
    scenarios where the observed excess corresponds to the
    next-to-lightest Higgs while the lightest one (with a mass $\sim $
    100 GeV) has not been detected
    yet~\cite{Heinemeyer:2011aa}. Although this possibility is
    appealing, and can certainly be accommodated in our model, here we
    restrict our analysis to the more conservative assumption that
    ATLAS and CMS collaborations have discovered the lightest
    (CP-even) Higgs eigenstate.}. As discussed in section
  \ref{subsec:sector} the two lightest eigenstates of the squared mass
  matrix in eq.~\eqref{matrix} are practically orthogonal to $\xi_0$
  for large enough $m_\Sigma^2$. They have masses
\be
m^2_{h,H}=\frac{1}{2}\left(\mathcal T\mp \Delta\right)
\label{autovalores}
\ee
with
\be
\mathcal T=\tr \mathcal M^2~,\qquad
\Delta=\sqrt{\mathcal T^2-4\mathcal D}~,\qquad
\mathcal D= \det \mathcal M^2~,
\ee
and they are related to the original fields as
\be
\left (\begin{array}{c} h_2\\h_1\end{array}
\right)=\left(\begin{array}{cc} \cos\alpha&\sin\alpha\\-\sin\alpha& \cos\alpha\end{array}
\right)
\left(\begin{array}{c} h\\H\end{array}
\right)
\label{eigenstates}
\ee
where the mixing angle $\alpha$ is determined by
\begin{equation}
\sin 2\alpha=\frac{2\mathcal M_{12}^2}{\Delta}\,,\quad
\cos 2\alpha=\frac{\mathcal M_{22}^2-\mathcal M_{11}^2}{\Delta}~.
\label{sina}
\end{equation}

\begin{figure}[htb]
\begin{center}
\vspace{5mm}
\includegraphics[width=80mm,]{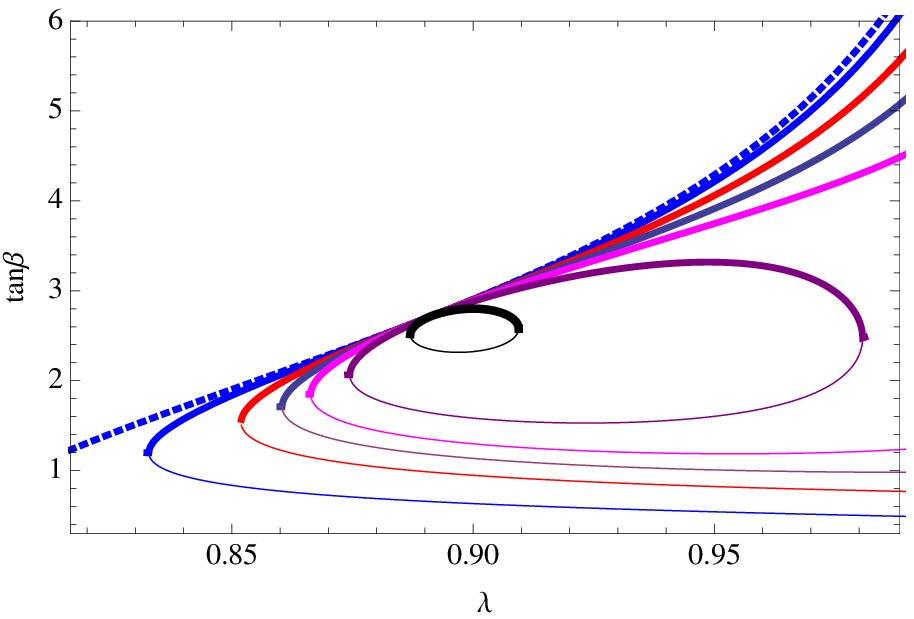}
\includegraphics[width=80mm,]{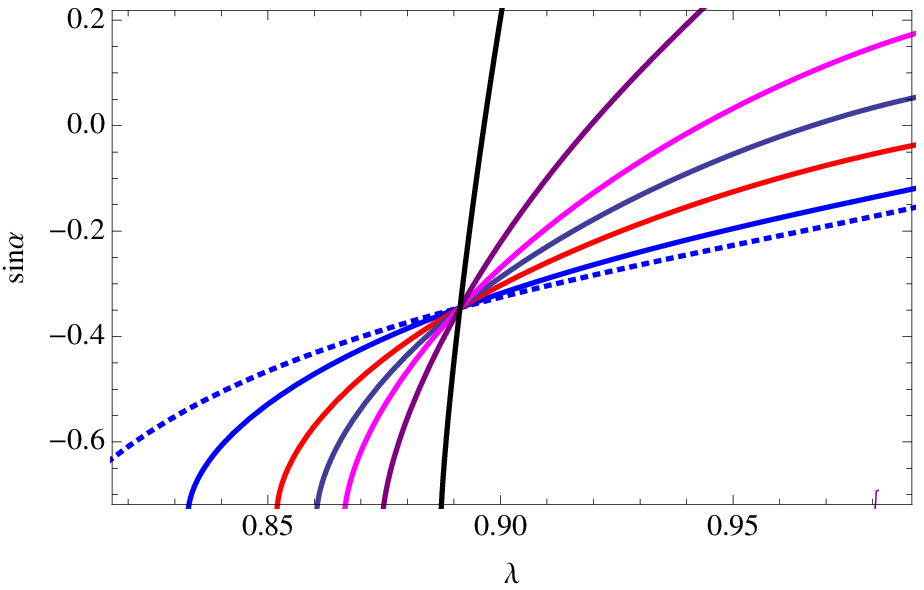}
\end{center}
\caption{\it Left panel: $\tan\beta$ as a function of $\lambda$
  providing $m_h=126\,$GeV in the decoupling limit or large $m_A$
  (blue dotted) and for $m_A=$ 200 GeV (blue solid), 155 GeV (red
  solid), 145 GeV (grey solid), 140 GeV (magenta solid), 135 GeV
  (purple solid) and 130 GeV (black solid). The other parameter inputs
  are those of eq.~\eqref{setting}. Right panel: The same but for
  $\sin\alpha$ as a function of $\lambda$.}
\label{fig:1}
\end{figure}

Once one fixes $m_A$, the constraint $m_h=126$\,GeV in
eq.~\eqref{autovalores} provides a relation between $\tan\beta$ and
$\lambda$. The curves satisfying this relation are plotted in
Fig.~\ref{fig:1} (left panel) for several fixed values of $m_A$ [the
above stop and triplet radiative corrections are included in
  $\mathcal M$ and evaluated for the parameter
setting~\eqref{setting}]. The function $\beta=\beta(\lambda;m_A)$ is
bivalued and we represent with a thick (thin) line the solution branch
corresponding to larger (smaller) value of $\tan\beta$.  We will use
this function to implement the condition $m_h=126$ GeV in the
following observables, and we will plot them with thick (thin) lines
when they correspond to the solution branch with largeer (smaller)
values of $\tan\beta$.

Independently of $m_A$, all lines intersect at the point
\be
\tan\beta_c\simeq 2.72~,\quad \lambda_c\simeq 0.89~.
\label{critical}
\ee
At such point the rotation angle diagonalizing $\M^2$ is independent
of $m_A^2$ (see Fig.~\ref{fig:1}, right panel). Since this point (also)
belongs to the line $m_A\to\infty$, the usual decoupling-limit
relation $\alpha_c=\beta_c-\pi/2$ is satisfied at small $m_A$ as well,
and the tree-level couplings of the Higgs $h$ to the SM particles are
hence those of the SM. For this reason the parameter region around
($\tan\beta_c,\lambda_c$) might provide a phenomenology very similar
to the one of the SM (as discussed in the next sections, relevant
differences may consist in the $h$ diphoton and invisible channels,
and in the $A,H^0,H^\pm$ decays if $m_A$ is light). 

We can understand the origin of the {\it intersection point}
\eqref{critical} as follows. The function $\beta=\beta(\lambda; m_A)$ in
the left panel of Fig.~\ref{fig:1} is the solution of the following
equation [cf.~also eq.~(\ref{autovalores})] for a fixed value of $m_h$, in particular for $m_h=126$ GeV:
\be
\mathcal D-m_h^2\mathcal T+m_h^4=0~.
\label{ligero}
\ee
Eq.~(\ref{ligero}) is a first order polynomial in $m_A^2$ (the
coefficient of $m_A^4$ cancels out) which can then be written as
\be
A(\tan\beta,\lambda)m_A^2+B(\tan\beta,\lambda)=0~.
\label{poly}
\ee 
For the point $(\tan\beta_c,\lambda_c)$ such that
\be
\begin{cases} A(\tan\beta_c,\lambda_c)=0 \\[1mm]
B(\tan\beta_c,\lambda_c)=0\end{cases}\label{sistema}
\ee
eq.~(\ref{poly}) [and (\ref{ligero})] is fulfilled for any $m_A$, and
any solution $\beta=\beta(\lambda;m_A)$ then crosses this point (for
the given $m_h$ and independently of $m_A$~\footnote{Had we included the (small) radiative corrections stemming from the Higgs sector the solution to eq.~(\ref{sistema}) $(\tan\beta_c,\lambda_c)$ would exhibit a tiny inappreciable (logarithmic) dependence on $m_A$ which we are neglecting throughout this paper.}).

It is straightforward to find the analytical solution of the system
(\ref{sistema}) for the tree-level squared mass $\M^2_0$. It is given
by $\tan\beta_c^0=1$ and $\lambda_c^0=\sqrt{2} m_h/v\simeq 1.02$, and
one can easily check that $\alpha_c^0=\beta_c^0-\pi/2$ diagonalizes
$\M^2_0$.  In the presence of radiative corrections the analytic
solution is more cumbersome. It can be found in the Appendix and
yields the critical values quoted in eq.~(\ref{critical}). As the
  analytic expressions show, the dependence of $\tan\beta_c$ and
  $\lambda_c$ on stop and triplet mass parameters is mild due to
  logarithmic suppressions and, in general, the larger these mass
  parameters the lower $\lambda_c$ and the higher $\tan\beta_c$ 
  (see Section~\ref{sec:6} for a numerical example which in
  particular yields $0.85\lesssim\lambda_c\lesssim 1$ and
    $1.5\lesssim \tan\beta_c\lesssim 3$ as typical values). Let us
  finally comment that Eq.~(\ref{sistema}) does not have a solution in
  the MSSM. On the contrary, in theories with an extra quartic
  coupling in the tree-level potential, as the present model
  containing the additional triplet or even in theories with extra
  singlets, there can be in general a solution which implies the
  existence of the discussed SM-like intersection point.

\section{The Higgs couplings}
\label{sec:coupl}
The angle $\alpha$ plays a fundamental role in the interactions of the
CP-even Higgses with the SM fields, as shown in Table~\ref{tabla}
where the ratios $r^0_{\mathcal H XX}$ are the tree-level
approximation of the quantities~\cite{Carena:1995bx}
\be
r_{\mathcal H XX}=\frac{g_{\mathcal H XX}}{g_{h XX}^{\rm SM}}\quad
{\rm with}\quad
\mathcal H=h,H; \quad X=W,Z,t,b,\tau,
\label{ratios}
\ee
being $g_{\mathcal H XX}$ and $g_{h XX}^{SM}$ the (effective)
couplings in the present theory and the SM, respectively.

\begin{figure}[htb]
\begin{center}
\vspace{5mm}
\includegraphics[width=80mm,]{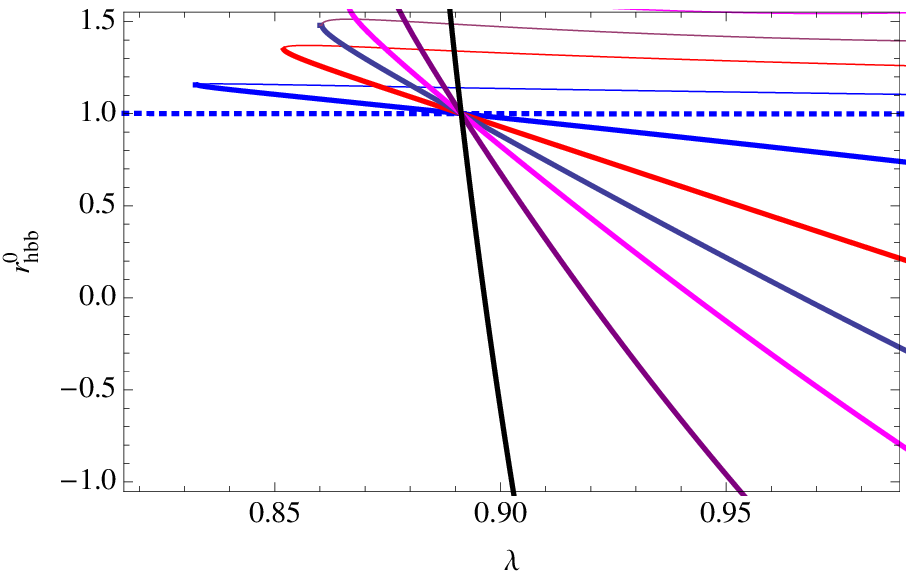}
\includegraphics[width=80mm,]{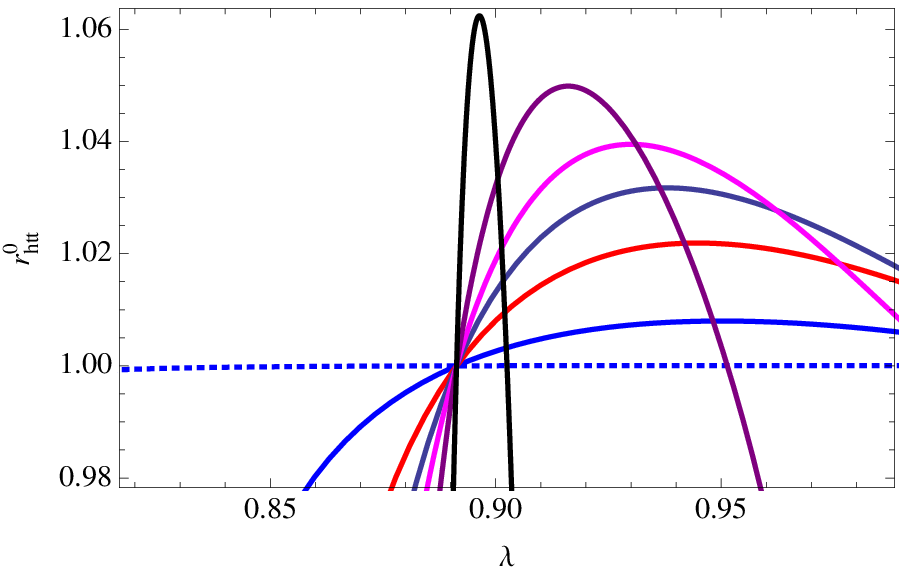}

\includegraphics[width=80mm,]{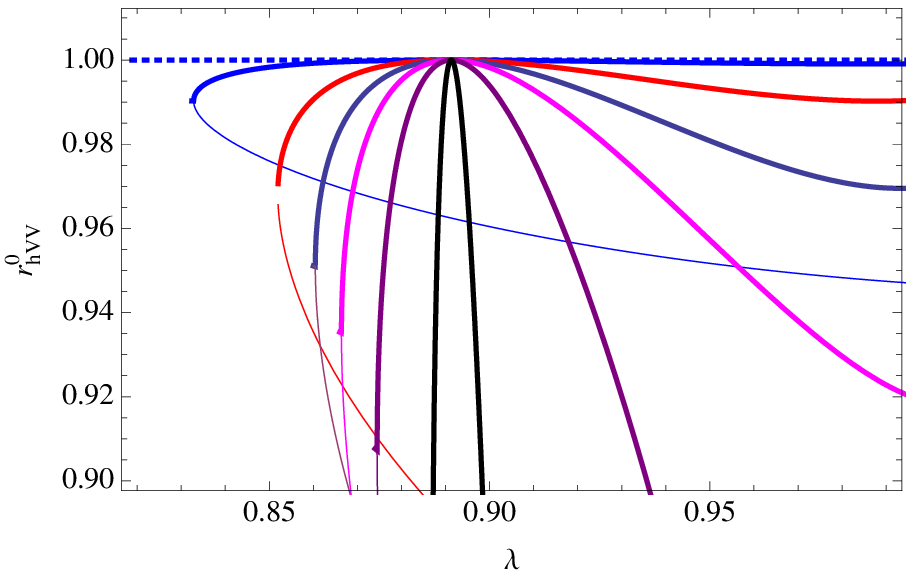}
\includegraphics[width=80mm,]{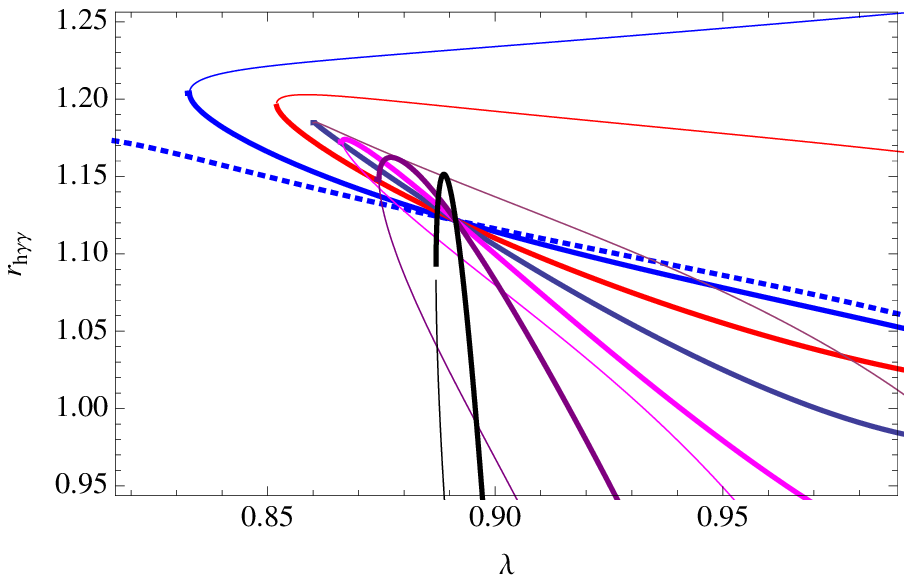}
\end{center}
\caption{\it Plot of the normalized tree-level couplings $r^0_{hbb}$
  (upper left panel), $r^0_{htt}$ (upper right panel) and $r^0_{hVV}$
  (lower left panel) and one-loop effective coupling
  $r_{h\gamma\gamma}$ (lower right panel) as a function of $\lambda$
  for different values of $m_A$. The constraints $m_h=126\,$GeV,
  $m_{\chi_1^\pm}=104\,$GeV and setting \eqref{setting} are assumed.
  Thick (thin) lines correspond to the large (small) $\tan\beta$
  branch in Fig.~\ref{fig:1}. Colour code is the same as in
  Fig.~\ref{fig:1}.}
\label{fig:2}
\end{figure}

The tree-level ratios $r_{htt}^0$, $r_{hbb}^0=r^0_{h\tau\tau}$ and
$r_{hVV}^0\equiv r_{hWW}^0=r_{hZZ}^0$ are plotted in Fig.~\ref{fig:2}
for fixed values of $m_A$ and along the curves
\begin{table}[htb]
\begin{center}
\begin{tabular}{||c|c|c|c|c|c||}
\hline
\rule{0pt}{4mm}
$r_{hWW}^0=r_{hZZ}^0$&$r_{HWW}^0=r_{HZZ}^0$&$r_{htt}^0$&$r_{Htt}^0$&$r_{hbb}^0=r_{h\tau\tau}^0$&$r_{Hbb}^0=r_{H\tau\tau}^0$\\[1mm]
\hhline{======}
\rule{0pt}{3.5ex}    
$\sin(\beta-\alpha)$&$\cos(\beta-\alpha)$&${\displaystyle \frac{\cos\alpha}{\sin\beta} }$&${\displaystyle \frac{\sin\alpha}{\sin\beta} }$&${\displaystyle -\frac{\sin\alpha}{\cos\beta}}$&${\displaystyle \frac{\cos\alpha}{\cos\beta} }$\\[2mm]
\hline    
\end{tabular}
\end{center}
\caption{\it The tree-level value of ratios (\ref{ratios}) for the different channels.}
\label{tabla}
\end{table}
$\beta=\beta(\lambda;m_A)$ of Fig.~\ref{fig:1}. As explained above,
all of them are equal to one at the intersection value
$(\tan\beta_c,\lambda_c)$. In particular this is so for $r_{hbb}^0$
(Fig.~\ref{fig:2} left upper panel). However moving away from the
intersection point one can easily modify $r_{hbb}^0$ by $\mathcal
O(\pm 1)$ factors, depending on the values of $m_A$.  On the other
hand for either sufficiently large $m_A$ (decoupling region) or for
parameter points near the SM-like intersection point
$(\tan\beta_c,\lambda_c)$, the $bb$ and $\tau\tau$ production via
Higgs decays, which respectively behave like $(r_{hbb}^0)^2$ and
$(r_{h\tau\tau}^0)^2$ at leading order, may appear in very good
agreement with the SM predictions, provided that the other couplings
take values close enough to the SM ones (more details will be given in
section~\ref{sec:4}).

Radiative corrections can of course induce important modifications to
the tree-level couplings. In particular, because of sbottom-gluino
(stop-gluino) loops, the ratio $r^0_{hbb}$ ($r^0_{htt}$) gets
renormalized as~\cite{Carena:1998gk}
\begin{eqnarray}
r_{hbb}&=&r_{hbb}^0 \left[1-\frac{\Delta(m_b)}{1+\Delta(m_b)}\left(1+\frac{1}{\tan\alpha\tan\beta}  \right)\right]~, \nonumber\\
r_{htt}&=&r_{htt}^0 \left[1-\frac{\Delta(m_t)}{1+\Delta(m_t)}\left(1+\tan\alpha\tan\beta  \right)\right] ~,
\label{hbb_rad}
\end{eqnarray}
with
\begin{eqnarray}
\Delta(m_b)&=&\frac{2\alpha_3}{3\pi}\tan\beta\, M_3\mu f(m_{\tilde b},M_3)~,\nonumber\\
\Delta(m_t)&=&\frac{2\alpha_3}{3\pi\tan\beta}\, M_3\mu f(m_{\tilde t},M_3)~,\nonumber\\
f(x,y)&=&\frac{x^2-y^2+y^2\log y^2/x^2}{(x^2-y^2)^2}~,
\end{eqnarray}
where $M_3$ and $m_{\tilde b}$ are respectively the gluino and sbottom
masses (in the sbottom sector we consider $m_D=m_Q$ and $A_b=0$).  The
couplings $g_{h tt}$ and $g_{h bb}$ may then depart from the
corresponding SM values in the presence of rather light gluinos and
third generation squarks. The same effect does not instead arise for
the decays involving the $\tau$ leptons and first two generation quarks
because of the small $\tan\beta$ regime imposed by the requirement
$m_h=126\,$GeV (cf.~left panel of Fig.~\ref{fig:1}). In Higgs search
data these radiative corrections might be an important signature to
discriminate experimentally the SM from the present triplet scenario
if its parameters are close enough to the intersection point, the
remaining Higgses cannot be detected and charginos are relatively
heavy (i.e.~if there is no sizeable Higgs diphoton enhancement, as we
will discuss next).

Radiative corrections are also crucial for loop-induced
decays. Charginos can provide sizeable contributions to $g_{\mathcal H
  \gamma\gamma}$ in addition to those of the top and $W$ boson already
present in the SM. Because of the electrically-charged triplet
fermions, the effective coupling $g_{h \gamma\gamma}$ can be enhanced
much more than in the MSSM~\footnote{ For an earlier analysis see~Ref.~\cite{DiChiara:2008rg}.}, even though it does
not differ from the SM coupling if charginos are heavy or from
  the MSSM one if $\mu_\Sigma$ is large and $\lambda$ is small.
The consequent increasing of $\Gamma(h\to\gamma\gamma)$ has been
already studied in the decoupling limit~\cite{Delgado:2012sm} and here
we extend that analysis to more general cases.

The chargino sector already contained in the MSSM $(\tilde
W^{\pm},\tilde H_1^-,\tilde H_2^+)$ mixes with the triplet charginos
$(\tilde\xi_1^-,\tilde\xi_2^+)$. Their mass matrix is given by
\be
\left(\tilde W^-,\tilde H_1^-,\tilde\xi_1^-\right)\mathcal M_{1/2}^\pm \left(\begin{array}{c}
\tilde W^+\\ \tilde H_2^+\\ \tilde\xi_2^+
\end{array}
\right),\quad \mathcal M_{1/2}^\pm=\left(\begin{array}{ccc}
M_2&gv_2&0\\
gv_1& \mu & -\lambda v_2\\
0& -\lambda v_1& \mu_\Sigma
\end{array}
\right)~,
\ee
and their contribution to $g_{\mathcal H \gamma\gamma}$ can be
determined from the QED effective potential~\cite{Ellis:1975ap,Shifman:1979eb,Carena:2012xa}
\be
\mathcal
L^{1/2}_{\gamma\gamma}=F_{\mu\nu}^2\frac{\alpha}{16\pi}2b_{1/2} 
\log\det \mathcal M_{1/2}^{\pm}(v_i+h_i/\sqrt{2})~,
\label{effective}
\ee
with $b_{1/2}=4/3$. By expanding $\mathcal L^{1/2}_{\gamma\gamma}$ to
linear order in $h_i$ and projecting onto the Higgs eigenstates $h,H$
one obtains
\begin{eqnarray}
r_{h\gamma\gamma}&=&
\frac{A_1(\tau_W) \sin(\beta-\alpha)+b_{1/2}A_{1/2}(\tau_t){\displaystyle \frac{\cos\alpha}{\sin\beta}}+
{\displaystyle\frac{b_{1/2}\cos(\alpha+\beta)v^2 (M_2\lambda^2+g^2\mu_\Sigma)}{\sin\beta\cos\beta v^2(M_2\lambda^2+g^2\mu_\Sigma)-M_2\mu\mu_\Sigma} }
} {A_1(\tau_W)+b_{1/2}A_{1/2}(\tau_t)}
\label{gammah}\nonumber\\[2mm] 
r_{H\gamma\gamma}&=&
\frac{A_1(\tau_W) \cos(\beta-\alpha)+b_{1/2}A_{1/2}(\tau_t){\displaystyle \frac{\sin\alpha}{\sin\beta}}+
{\displaystyle\frac{b_{1/2}\sin(\alpha+\beta)v^2 (M_2\lambda^2+g^2\mu_\Sigma)}{\sin\beta\cos\beta v^2(M_2\lambda^2+g^2\mu_\Sigma)-M_2\mu\mu_\Sigma} }
} {A_1(\tau_W)+b_{1/2}A_{1/2}(\tau_t)}
\nonumber \\
&&
\label{rhgammagamma}
\end{eqnarray}
where $A_1(\tau_W)\simeq-8.3$ and $A_{1/2}(\tau_t)\simeq 1.4$.
Eq.~(\ref{gammah}), which reproduces the result of
Ref.~\cite{Delgado:2012sm} in the decoupling limit, shows that
large enhancements can arise due to charginos. The exact amount of
increase depends on many parameters which might be constrained by
future searches at the LHC. For instance, in order to fulfill the
chargino mass bound $m_{\tilde\chi_1^\pm}\gtrsim 104$\,GeV~\cite{pdg}
with any sensible value of $\lambda$, one has to choose $|\mu|$ and
$|\mu_\Sigma|$ similar to (or larger than) those we are considering in
this analysis.  In the following we impose $m_{\tilde\chi_1^\pm}=
104$\,GeV which, in turn, determines the value of $M_2$ along the
curves $\beta=\beta(\lambda;m_A)$ of Fig.~\ref{fig:1}. In particular it
implies $M_2(\tan\beta_c,\lambda_c)\simeq 164$\,GeV for our setting
\eqref{setting}. Moreover at the intersection point \eqref{critical}
$r_{h\gamma\gamma}$ has to coincide for all the curves $\beta(\lambda;
m_A)$, independently of $m_A$, as a glance at 
expression~\eqref{rhgammagamma} suggests and Fig.~\ref{fig:2} (right
lower panel) clearly confirms. Notice that at the intersection point,
which depends on our parameter choice, the enhancement is
$r_{h\gamma\gamma}(\tan\beta_c,\lambda_c) \simeq 1.12$, but for smaller $m_\Sigma$ (larger $|\mu|$ and $|\mu_\Sigma|$) a
larger (smaller) $r_{h\gamma\gamma}$ at the new intersection point is
possible. More details on this issue will be presented in
section~\ref{sec:6}.

\section{Higgs production rates at the LHC}
\label{sec:4}
From the values of $r_{\mathcal HXX}$ determined in the previous
section one can compute the predicted signal strength $\mathcal
R_{\mathcal H XX}$ of the decay channel $\mathcal H\to XX$, with
$\mathcal H=h,H$ and $X=W,Z,t,b,\tau$:
%
%
%
\be
\mathcal R_{\mathcal H XX}=\frac{\sigma(pp\to \mathcal H) BR(\mathcal H\to XX)}{\quad\left[\sigma(pp\to h)BR(h\to XX)\right]_{SM}}~.
\ee
In particular for the gluon-fusion (ggF), the associated
production with heavy quarks ($\mathcal H tt$), the associated
production with vector bosons ($V\mathcal H$) and the vector boson
fusion (VBF) production processes, one can write
\begin{eqnarray}
\mathcal R_{\mathcal HXX}^{(ggF)}&=&\mathcal R_{\mathcal HXX}^{(\mathcal Htt)}=
\frac{r_{\mathcal Htt}^2\ r_{\mathcal HXX}^2}{\mathcal D}~,\qquad\qquad
\mathcal R_{\mathcal HXX}^{(VBF)}=\mathcal R_{\mathcal
  HXX}^{(V\mathcal H)}=\frac{r_{\mathcal HWW}^2\ r_{\mathcal
    HXX}^2}{\mathcal D}~,\nonumber\\
\mathcal D &=&
BR(h\to b\ b)_{SM} \ r_{\mathcal Hbb}^2 \ + \ BR(h\to gg,cc)_{SM} \ r_{\mathcal H tt}^2\nn\\
&+&BR(h\to \tau\tau)_{SM}\ r_{\mathcal H\tau\tau}^2 \;
+ BR(h\to WW,ZZ)_{SM}\ r_{\mathcal HWW}^2 ~,
%
\label{R}
\end{eqnarray}
\begin{figure}[htb]
\begin{center}
\vspace{5mm}
\includegraphics[width=80mm,]{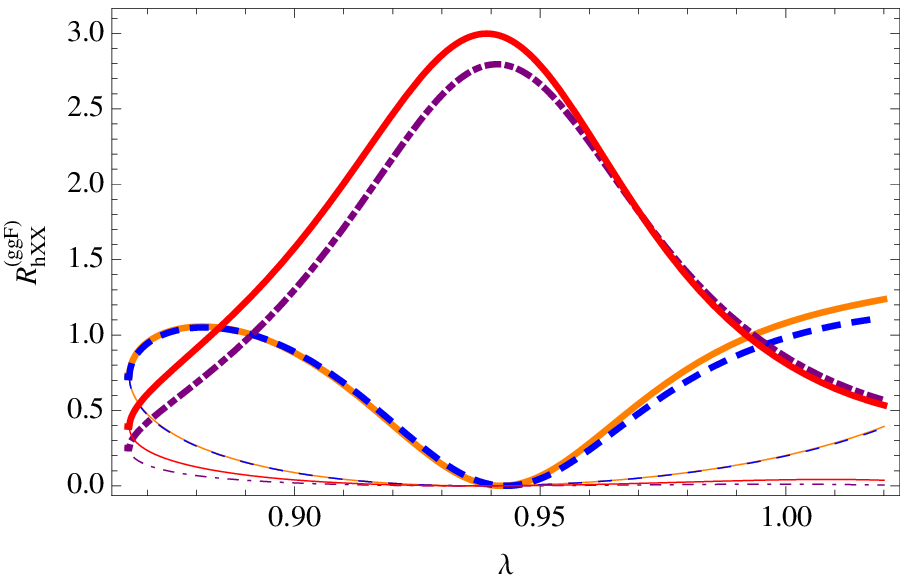}
\includegraphics[width=80mm,]{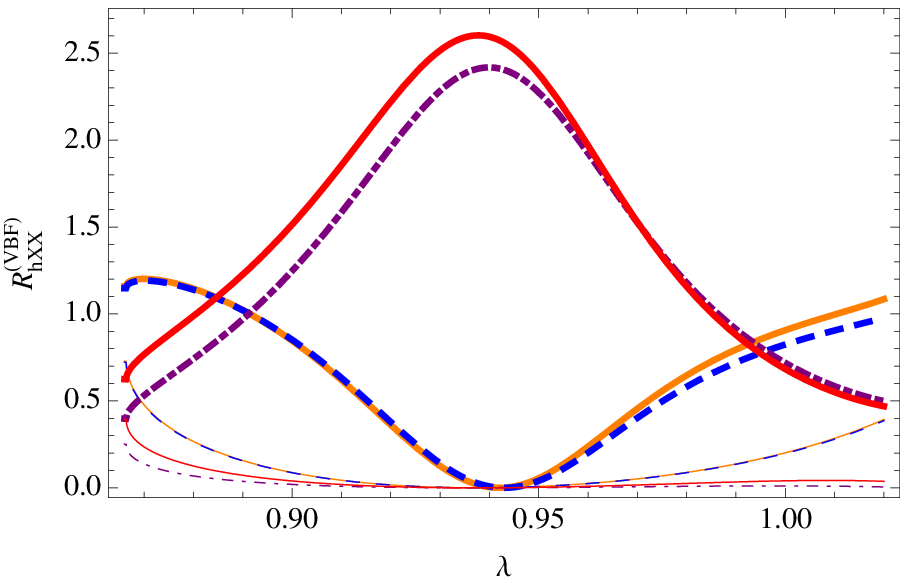}
\end{center}
\caption{\it Left panel: Signal strengths for the lightest Higgs
  production by the gluon-fusion mechanism followed by decay in
  different channels: $\mathcal R_{hbb}^{(ggF)}$ (solid orange line),
  $\mathcal R_{h\tau\tau}$ (dashed blue line), $\mathcal R_{hWW}$ (dot
  dashed purple line) and $\mathcal R_{h\gamma\gamma}$ (solid red
  line), for $m_A=140\,$GeV. The constraints $m_h=126\,$GeV,
  $m_{\chi_1^\pm}=104\,$GeV and setting~\eqref{setting} are
  implemented. Thick (thin) lines correspond to the large (small)
  $\tan\beta$ branch in Fig.~\ref{fig:1}. Right panel: The same but
  for the vector boson fusion production.}
\label{fig:3}
\end{figure}
where we assume that: {\it i)} $\Gamma(\mathcal H\to gg)$ is dominated
by the top loop so that $r_{\mathcal Hgg}=r_{\mathcal
  Htt}$~\footnote{However, when $r_{\mathcal H tt}\ll 1$, as it can be
  the case for the Higgs $H$, we consider also the correction from the
  bottom loop and then $r_{H gg}^2=\left[r_{Htt}+(m_b/m_t) r_{H
      bb}\right]^2.$}, which requires for instance that there is no
light colored supersymmetric particle (e.g.~light stop) strongly
coupled to the Higgs; {\it ii)} the invisible decay width of $\mathcal
H$ into neutralinos is negligible~\footnote{For a study where both
  assumptions are violated and the interplay between light stops and
  neutralinos is crucial to reproduce the LHC data, see
  e.g. Ref.~\cite{Carena:2012np}.}.

Besides the Higgs diphoton width, in the decoupling limit all
branching ratios of $h$ are in quite good agreement with those of the
SM (unless there are rare large radiative corrections in $bb$ and $tt$
channels), which are $BR(h\to bb)_{SM}=0.56$, $BR(h\to
\tau\tau)_{SM}=0.06$, $BR(h\to WW,ZZ)_{SM}=0.26$, $BR(h\to
gg,cc)_{SM}=0.11$ at $m_h=126\,$GeV (we also take $\Gamma_{SM}=4.2$
MeV)~\cite{Denner:2011mq}. The phenomenology of this case has been
studied in Ref.~\cite{Delgado:2012sm}~\footnote{Some numerical
  discrepancies may be detected between the analysis of
  Ref.~\cite{Delgado:2012sm} and the present one which are due to the
  use of different inputs.} for large values of $m_A$. In this paper
we instead focus on the regime of small $m_A$. As discussed below, the
presence of light $A$, $H$ and $H^{\pm}$ might be the key signature to
distinguish the triplet scenario from the SM at the LHC.

For $m_A=140$ GeV the behavior of the signal strengths $\mathcal
R_{hXX}$ along the curves $\beta(\lambda; m_A)$ of Fig.~\ref{fig:1}
are exhibited in Fig.~\ref{fig:3}. The decays originated by gluon
(vector boson) fusion Higgs production are presented in the left
(right) panel of the figure. As we can see, only thick lines, which
correspond to the larger-$\tan\beta$ branch of solutions in
Fig.~\ref{fig:1}, are phenomenologically relevant if one requires the
LHC observed excess to be related only to the lightest Higgs, as we do
in this paper.  We have fixed the relative signs of $\mu$ and $M_3$
such that $sign(\mu M_3)<0$. Because of this assignment, $bb$ versus
$\tau\tau$ production is enhanced at $\lambda\gtrsim 0.95$. Had we
chosen $sign(\mu M_3)>0$ the $bb$ (solid orange) curve would underlie
the $\tau\tau$ (blue dashed) line. On the other hand, different
choices of $M_3$ and $m_{\tilde b}$ could have slightly increased or
largely suppressed the relative enhancement with respect to the case
plotted in Fig.~\ref{fig:3}. Moreover notice that $\Delta(m_b)$
generates a relatively small splitting between $bb$ and $\tau\tau$
channels, especially at the intersection point \eqref{critical},
indicating \textit{a posteriori} that radiative corrections which are
subleading to $\Delta(m_b)$ are negligible, as we are assuming in this
analysis.

An important issue highlighted in Fig.~\ref{fig:3} is the possibility
of having SM-like production and decay rates in the small-$m_A$
regime. At the intersection point (\ref{critical}) {\it the signal
  strengths of $bb,\tau\tau,WW$ and $ZZ$ channels are as in the SM}
(unless of subleading corrections that can be modulated by different
choices of the spectrum), with the notable difference of an about
$25\%$ {\it excess in the diphoton Higgs decay}, which could be
reduced for larger values of chargino masses or increased for smaller
values of $m_\Sigma$, as discussed in section~\ref{sec:6}. Therefore
values of $(\tan\beta,\lambda)$ near $(\tan\beta_c,\lambda_c)$, as
well as in the peculiar region around $(\tan\beta,\lambda)=
(4.6,0.995)$ of Fig.~\ref{fig:3}, seem the most promising to adjust
future experimental data once their statistical and systematic
uncertainties had been reduced~\footnote{We wish to stress here that
  the existence of the SM-like intersection point does not rely on the
  chosen values of the parameters, and different settings from that in
  eq.~\eqref{setting} would only influence the values of
  $(\tan\beta_c,\lambda_c)$ and consequently the amount of diphoton
  excess. The second peculiar point for $\lambda\simeq 1$ is instead
  strongly parameter dependent and only exists for small values of
  $m_A$.}.

The main feature of the peculiar region arising around $\lambda=
0.995$ for $m_A=140\,$GeV (it also exists for other small values of
$m_A$) is that $\tan\beta$ is relatively large and
$\sin\alpha/\!\cos\beta\!=\!\mathcal O(1)$ (cf.~Figs.~\ref{fig:1} and
\ref{fig:2}). The production of $WW,ZZ$ therefore reproduces the one
predicted in the SM. The diphoton signal strength is also SM-like
because the enhancement due to charginos is suppressed by the
relatively large value of $\tan\beta$ yielding small
$\cos(\alpha+\beta)$ ($\alpha$ is positive and $\tan\alpha\simeq
1/\tan\beta $). Finally the production of $bb$ is enhanced with
respect to the $\tau\tau$ production because the radiative correction
$\Delta(m_b)$ is proportional to $\tan\beta$. However, as stressed
above, this relative enhancement can be reversed by changing
$sign(M_3\mu)$ or can be made tiny for other values of $m_{\tilde b},
\mu$ and $M_3$. In any case in this parameter region the model
predicts $h\to\tau\tau$ rates that tend to be suppressed with respect
to the SM in every Higgs production channel.

A last remark on the region with large values of $\lambda$ (namely
$\lambda\gtrsim 0.95$ for $m_A=140\,$GeV): the signs of $g_{hbb}$ and
$g_{h\tau\tau}$ are opposite to those in the SM but their absolute
values are close to 1 [cf.~Fig.~\ref{fig:2}; notice also that the
small radiative correction in eq.~\eqref{hbb_rad} does not flip the
sign of $g_{hbb}$]. These unusual signs are not ruled out in the
regime $\tan\beta\ll m_t/m_b$ that emerges in Fig.~\ref{fig:1}. In
such a case, indeed, the absolute values of the couplings $g_{hbb}$
and $g_{h\tau\tau}$ are much smaller than the top-quark Yukawa
coupling, and interference in observables sensitive to either
$sign(g_{hbb})$ or $sign(g_{h\tau\tau})$ are difficult to detect
(unlike the case of $sign(g_{htt})$, see
e.g.~Ref.~\cite{Espinosa:2012ir}).

\begin{figure}[htb]
\begin{center}
\vspace{5mm}
\includegraphics[width=80mm,]{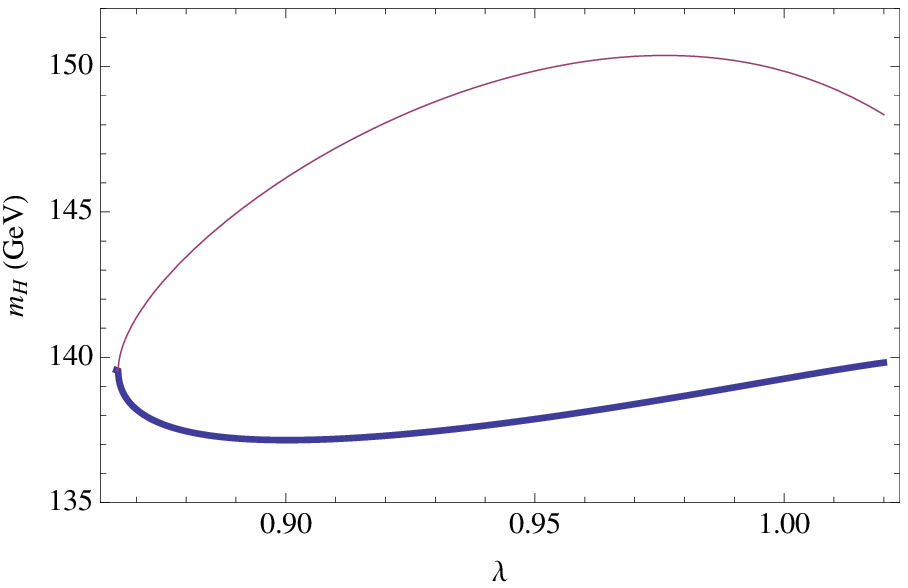}
\includegraphics[width=80mm,]{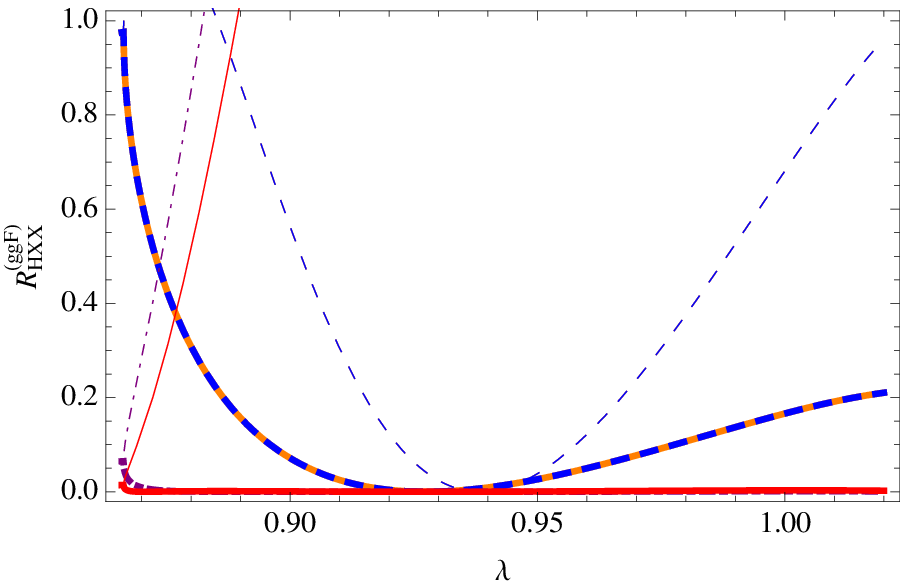}
\end{center}
\caption{\it Left panel: Plot of the next-to-lightest Higgs mass $m_H$
  as a function of $\lambda$ for $m_A=140\,$GeV. The constraint
  $m_h=126\,$GeV and setting~\eqref{setting} are implemented. Thick
  (thin) lines correspond to the large (small) $\tan\beta$ branch in
  Fig.~\ref{fig:1}.  Right panel: The same as in the left panel of
  Fig.~\ref{fig:3} but for the next-to-lightest Higgs signal strengths
  $\mathcal R_{HXX}^{(ggF)}$ produced by gluon fusion.}
\label{fig:4}
\end{figure}
On the other hand even in the SM-like intersection region with all
$\mathcal R_{hXX}$ equal to 1 (but $m_A$ small), one can try to
discriminate the SM from this scenario by looking at the remaining
Higgs fields whose masses are in the energy range probed by the
LHC. Focusing first on the field $H$, its mass $m_H$ as a function of
$\lambda$ at $m_A=140\,$GeV is shown in the left panel of
Fig.~\ref{fig:4}. The thickness code is the same as in the previous
figures. Since for the thin lines all signal strengths $\mathcal
R_{hXX}$ are very suppressed~\footnote{We remind the reader that we
  assume the field $h$ to be the recently observed particle and the
  unique Higgs in the range $123\,{\rm GeV}\lesssim m_h\lesssim
  126\,{\rm GeV}$. Other Higgs spectrum scenarios, such as for
  instance $m_h=123\,$GeV (or $m_h\ll 123\,$GeV) and $m_H=126\,$GeV,
  would require a dedicated analysis which is beyond the scope of the
  present paper.}, the phenomenologically relevant curves are the
thick ones, which correspond to $m_H\simeq 139$\,GeV. The field $H$
has therefore small signal strengths $\mathcal R_{HXX}$ at the values
of $\lambda$ providing realistic light-Higgs phenomenology
[cf.~Fig.~\ref{fig:3}]. In particular at the promising SM-like
intersection point $\lambda\simeq\lambda_c$ the productions of $WW$,
$ZZ$ and $\gamma\gamma$ are unobservable and there is only a little
fermion production with $\mathcal R_{H\tau\tau}\simeq \mathcal
R_{Hbb}\simeq 0.1$. Furthermore, if the lightest neutralino was in the
(short) mass range $m_h/2<m_{\tilde \chi_1^0}<m_H/2$ the invisible
channel $H\to \tilde\chi_1^0\tilde\chi_1^0$ could further dilute the
fermion branching ratios without altering the signal strengths
$\mathcal R_{hXX}$. Looking for signatures of light $A$ and $H^{\pm}$
seems therefore more promising.

Quantifying the $A$ and $H^{\pm}$ field phenomenology is strongly
parameter dependent and correlating it to the CP-even Higgs signatures
would require dedicated analyses that we leave for future
research. Nevertheless we can already envisage the main features since
the $A$ and $H^{\pm}$ are expected to have signatures quite similar to
those in the MSSM. Their couplings to the SM fields are indeed
obtained by rotating by the angle $\beta$ the CP-odd and charged
components of $H_{1,2}$\,, like in the MSSM. The relevant difference
is of course the mass spectrum. For instance, for the cases where we
can find $\mathcal R_{hXX}\approx 1$, the relation $m_{H^\pm}>m_t$
arises at even relatively small $m_A$, unlike in the MSSM and its
  singlet extensions [cf.~eq.~\eqref{chargedHiggs}].
It thus escapes from the experimental constrains which
  focus on the possibility of the charged Higgs being produced as a
  decay of the top. The branching ratios $BR(H^\pm\to tb)$ and
$BR(H^\pm\to \tau\nu_\tau)$ are then expected to be the dominant ones,
possibly along with $BR(H^\pm\to \chi_1^0\chi_1^\pm)$ when the channel
is kinematically accessible. Moreover, although strongly model
dependent, some relevant constraints on the $H^{\pm}$ radiative
corrections are possible, e.g.~the experimental measure of $B\to X_s
\gamma$ is expected to bound the $H^{\pm}$ versus $\chi_1^\pm$ space of
parameters.

Also the CP-odd Higgs signatures should be basically similar to those
in the MSSM. The field $A$ is produced only through gluon
  fusion and decays mainly into bottom pairs. Its tree-level coupling to weak
  vector bosons is zero and its decay into photons is two
  orders of magnitude smaller than the one of
  $h$~\cite{Djouadi:2005gj}.  Moreover for the case $m_A=140\,$GeV
that we have previously considered, the decay channel $A\to hZ$ is
kinematically forbidden (it opens up at $m_A\gtrsim 217\,$GeV).
Understanding whether the decay products of a light $A$ field appear
in the LHC Higgs search analyses as particles produced by the $h$
decay, and therefore whether our parameter regions around the
intersection points are actually compatible with the reinterpreted
experimental data for low $m_A$, would require a much more refined
study than the one we have carried out here. In this sense we stress
that the qualitative results we highlighted at $m_A=140\,$GeV, are
also valid at $m_A$ large enough not to introduce the above potential
subtlety.  On the other hand, the same problem would exist also in the
MSSM. For this reason we expect that the experimental bounds on $m_A$
and $m_H^\pm$ in the present model are similar to those obtained in
the MSSM~\footnote{Of course, only MSSM experimental bounds that do
  not make use of the constraint $m_h\approx 126\,$GeV in the analysis
  should apply. For updated studies see talks in
  Ref.~\cite{confHiggs}.} from direct searches. In particular since
CMS and ATLAS mainly focus on the bottom signal for the Higgs in
association with a weak vector boson~\cite{Atlas,CMS}, those bounds do
not apply to the CP-odd Higgs as the latter is not produced in that
channel.

\section{Accuracy of the decoupling approximation}
\label{sec:5}

The results of the previous sections are obtained by neglecting the
triplet-Higgs mixing terms in eq.~\eqref{matrix}. Here we check the
goodness of this approximation for our choice $m_\Sigma=5\,$TeV.
First of all let us remind that in all cases neglecting the VEV of the
neutral triplet field $\xi^0$ in the minimization and squared-mass
equations is a consequence of the experimental constraint on the
$T$-parameter which imposes $m_\Sigma\gtrsim 1.5$ TeV if no parameter
tuning is present. However for $m_\Sigma\sim 1.5$ TeV, although one
can certainly neglect the value of $\langle\xi^0\rangle$, one cannot
ignore the matrix elements
%
\begin{eqnarray}
\widehat{\mathcal M}_{31}^{\,2}=\widehat{\mathcal M}_{13}^{\,2}&=& (E_2\cos\beta+E_3\sin\beta)v\nonumber\\
\widehat{\mathcal M}_{32}^{\,2}=\widehat{\mathcal M}_{23}^{\,2}&=& (E_3\cos\beta+E_2\sin\beta)v
\label{offdiagonal}
\end{eqnarray}
where
\be
E_2=-\frac{\lambda}{\sqrt{2}}(A_\lambda+\mu_\Sigma),\quad
E_3=-\sqrt{2}\lambda\mu~,
\ee
which will modify the eigenvalues of the squared-mass matrix as well
\begin{figure}[htb]
\begin{center}
\includegraphics[width=80mm,height=60mm]{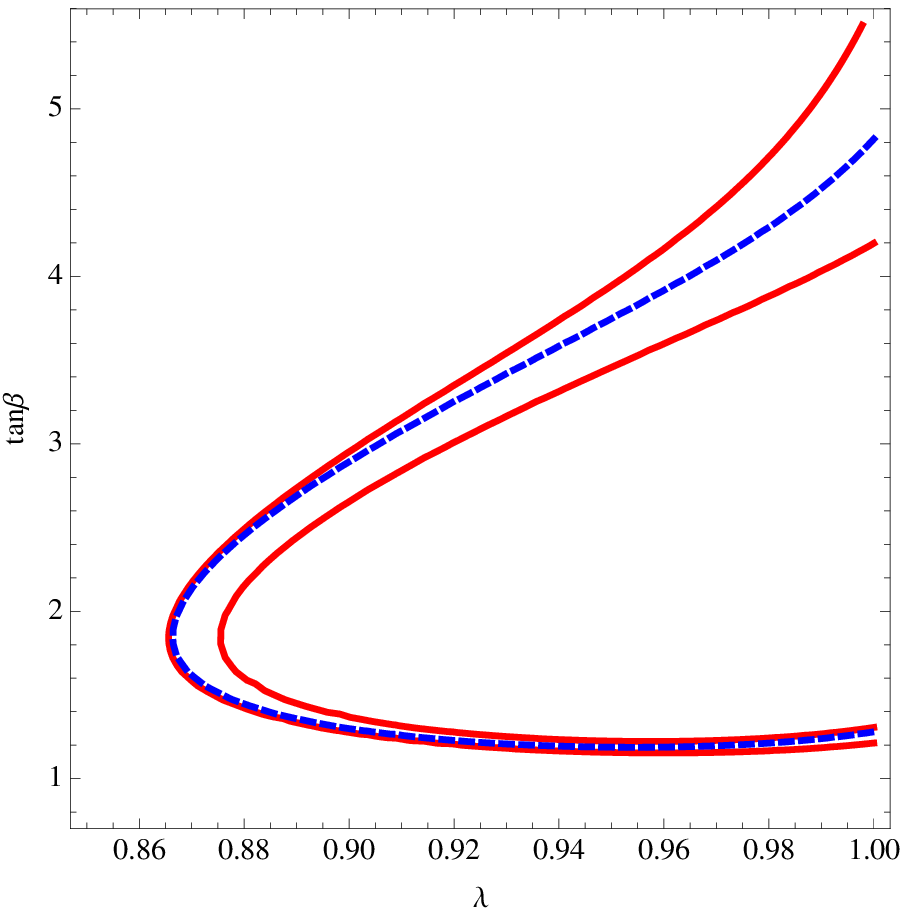}\hspace{.1cm}
\includegraphics[width=80mm,height=60mm]{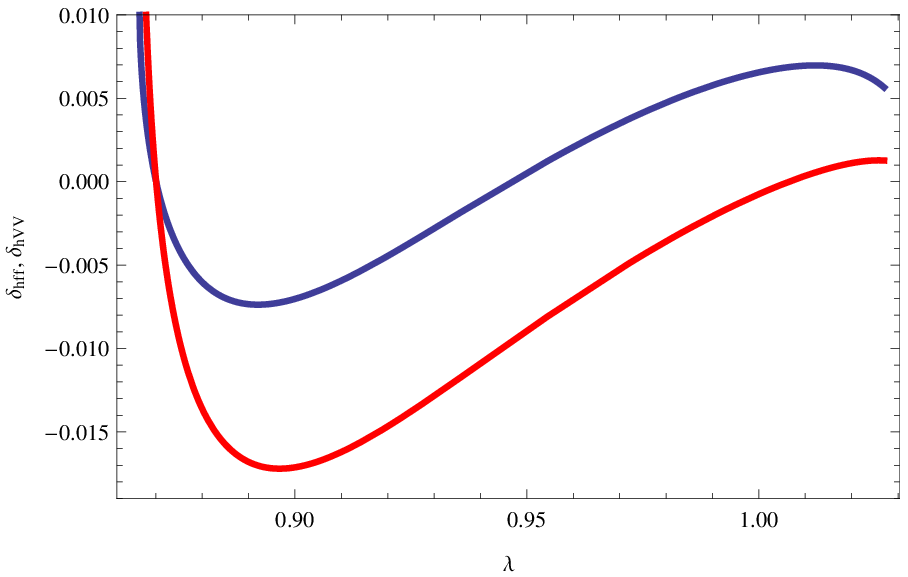}
\end{center}
\caption{\it Left panel: Contour plots in the $(\lambda,\tan\beta)$
  plane for $m_h=126$ GeV in the approximation of decoupling triplet
  scalars (blue dashed line) and $m_h=$125 GeV (outer red solid line) and 126 GeV (inner red solid line) in the exact theory. Right panel: Plots of $\delta_{hff}$
  (upper grey solid line) and $\delta_{hVV}$ (lower red solid line) as a function
  of $\lambda$ along the dashed contour of the left panel. }
\label{fig:5}
\end{figure}
as the rotation angles diagonalizing it, and thus the couplings of the
CP-even Higgs sector to $tt,\, bb,\, \tau\tau,\, WW,\, ZZ$. In other
words the experimental constraint on the $T$-parameter
($m_\Sigma\gtrsim 1.5$ TeV) is not strong enough to implement full
decoupling of the triplet scalars. Instead, for $m_\Sigma=5$ TeV the
decoupling approximation that neglects the entries \eqref{offdiagonal}
is good enough, as we now check.

In the left panel of Fig.~\ref{fig:5} we plot the constraint on
$\lambda$ and $\tan\beta$ coming from the requirement $m_h=126$ GeV in
the decoupling approximation (dashed blue line) with $m_A=140$ GeV and
the parameter setting \eqref{setting}. The curve lies in between the
lines corresponding to $m_h=125$\,GeV (outer solid red line) and
$m_h=126$\,GeV (inner solid red line) derived from the full
squared-mass matrix $\widehat{\mathcal M}^2$. We see that the decoupling
approximation at $m_\Sigma=5\,$TeV never overestimates the mass $m_h$
by more than 1\,GeV, which is well within the theoretical
uncertainties of our calculation of the lightest Higgs mass and the
experimental errors in the determination of the Higgs mass at the LHC.
In particular the overestimate at $\lambda=\lambda_c$ could have been
easily absorbed by shifting by about 1\,GeV the input value of $m_h$
in the approximated theory, i.e.~from $m_h=126$ GeV to $m_h=127$ GeV.

Besides the overestimate on the lightest Higgs mass, one also has to
check the error on the lightest-Higgs couplings.  Their uncertainties
are estimated in the right panel of Fig.~\ref{fig:5} where, considering
$\beta=\beta(\lambda)$ along the dashed line of the left panel
(i.e.~imposing $m_h=126$ GeV in the approximated theory), we plot the
quantities $\delta_{hff}$ ($f=b,\, t,\, \tau $) and $\delta_{hVV}$
($V=W,\, Z$) defined as
\be
\delta_{hXX}=\frac{r_{hXX}^{ex}-r_{hXX}^{ap}}{r_{hXX}^{ex}}~,
\label{deltas}
\ee
where the superscript \textit{ex} (\textit{ap}) refers to the exact
(decoupling approximation) results. In all the cases the approximated
approach provides couplings with less than 1\% error, which proves
that the results of the previous sections are reliable as
expected. Moreover we can see that $\delta_{hXX}=0$ at $\lambda\simeq
0.86$ near, but not quite coincident with, the intersection point
(\ref{critical}).  This means, as we will see in detail in the next
section, that the existence of the (SM-like) intersection point is not
an artifact of the decoupling approximation [where
$r^{ap}_{hXX}(\lambda_c^{ap})=1$] but also appears in the
non-decoupled cases~\footnote{This fact can be easily related to the
  cancellation of the $m_A^4$ term in eq.~\eqref{ligero} which in turn
  implies that $m_h^2$ is not controlled by supersymmetry breaking
  parameters in the decoupling limit of the most general
  supersymmetric theory, but by the electroweak breaking scale $v^2$.}
[where $r^{ex}_{hXX}(\lambda_c^{ex})=1$] although its localization is
slightly shifted ($\lambda_c^{ex}\neq\lambda_c^{ap}$).

\section{Some comments on the non-decoupling regime}
\label{sec:6}
As we have previously noticed the parameter $m_\Sigma$ can be lowered
down to values around $1.5$ TeV, consistently with electroweak
experimental observables. However in the low $m_\Sigma$ region allowed
by the experimental constraint on the $T$ parameter, neglecting the
off-diagonal matrix entries (\ref{offdiagonal}) is a rude
approximation, as one can explicitly check. One then should study the
model by a full numerical analysis which is outside the scope of the
present paper.  Instead we can easily investigate the low-$m_\Sigma$
scenario at the (phenomenologically interesting) intersection points
fixing the Higgs mass $m_h=126$ GeV in the exact theory. At the
intersection point, indeed, the lightest-Higgs production and decay
rates are at tree-level those of the SM, and the effect of moving
\begin{figure}[htb]
\begin{center}
\includegraphics[width=80mm,height=64mm]{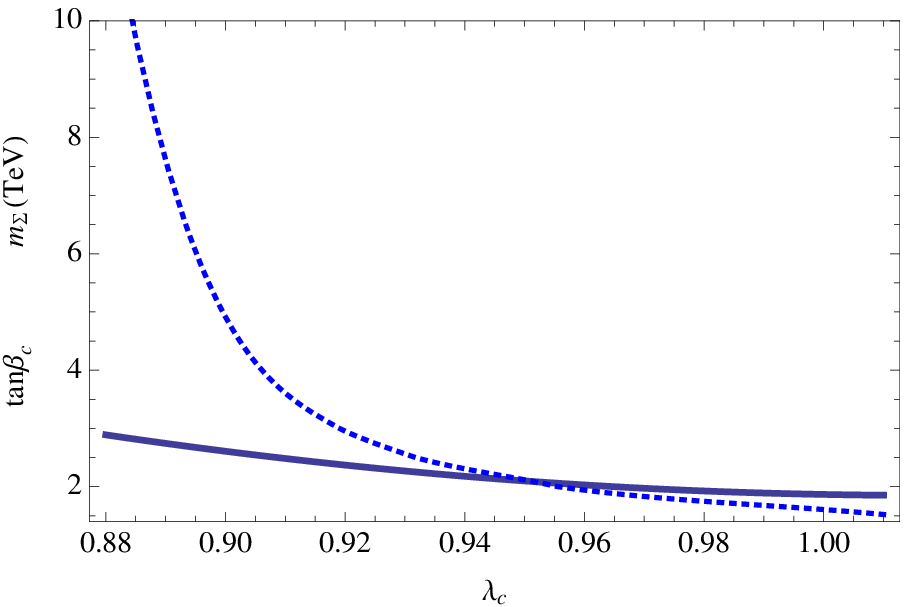}\hspace{.1cm}
\includegraphics[width=80mm,height=65mm]{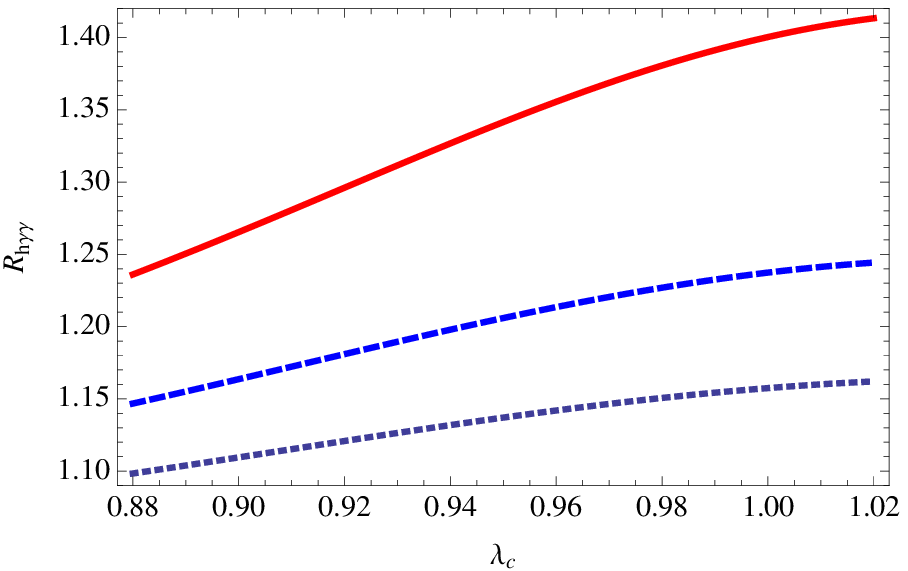}
\end{center}
\caption{\it Left panel: Values of $(\tan\beta_c,\lambda_c)$ (solid
  line) for $m_\Sigma$ in the range $1.5 \textrm{ TeV}\leq
  m_\Sigma\leq 10 \textrm{ TeV}$ (dotted line). The $126$\,GeV
  constraint on the lightest Higgs mass in the exact theory, the setting
  \eqref{setting} (except for $m_\Sigma$) is implemented.
  %
  %
  Right panel: The corresponding value of $\mathcal R_{h\gamma\gamma}$
  as a function of $\lambda_c$ along the intersection points for the
  different values of $m_\Sigma$ and the requirements
  $m_{\chi_1^\pm}=104\,$GeV (solid line),
  $m_{\chi_1^\pm}=\,150$ GeV with $\mu=\mu_\Sigma=\,300$ GeV (dashed line) and
  $m_{\chi_1^\pm}=\,200$ GeV with $\mu=\mu_\Sigma=\,350$ GeV (dotted line).}
\label{fig:6} 
\end{figure}
$m_\Sigma$ translates just in a modification of the actual value of
$(\lambda_c,\tan\beta_c)$ which the one-loop Higgs rates depend on
(there is no explicit dependence on $m_\Sigma$).

The position of the intersection point in the $(\lambda,\tan\beta)$
plane as a function of $m_\Sigma$ can be easily deduced from the left
panel of Fig.~\ref{fig:6}. For a given point of the $m_\Sigma$ curve
(dotted line), the point in the solid line having the same abscissa
gives the coordinates $(\lambda_c,\tan\beta_c)$ of the intersection
point in the scenario with that particular value of $m_\Sigma$.  As
already explained, departures of $\mathcal R_{hXX}$ from one (with $X$
a SM particle) appear at the intersection points only via loop
effects. In particular the Higgs diphoton channel is extremely
sensitive to the changes in $(\lambda_c,\tan\beta_c)$ caused by
modifications of $m_\Sigma$ [cf.~eq.~\eqref{rhgammagamma}]. This is
quantified in the right panel of Fig.~\ref{fig:6} where one can read
the value of $\mathcal R_{h\gamma\gamma}$ at the intersection point
for different values of $m_\Sigma$ by comparing the left and right
panels of the figure. For instance, looking at the solid curve of the
right panel (i.e.~$m_{\chi^\pm_1}=104\,$GeV with
$\mu=\mu_\Sigma=250\,$GeV), for $m_\Sigma=1.5\,$TeV the
$h\to\gamma\gamma$ decay rate is about 1.4 times larger than in SM at
the intersection point, which corresponds to $\lambda_c\simeq 1$ and
$\tan\beta_c\simeq 1.8$. However, the diphoton excess decreases by
raising the chargino spectrum. In particular at the intersection point
for $m_\Sigma=1.5\,$TeV it turns out that $\mathcal
R_{h\gamma\gamma}=$1.25 for $m_{\chi^\pm_1}=150\,$GeV, with
$\mu=\mu_\Sigma=300\,$GeV (dashed line), and $\mathcal R_{h\gamma\gamma}=$1.16
for $m_{\chi^\pm_1}=200\,$GeV, with $\mu=\mu_\Sigma=350\,$GeV (dotted
line). As expected, in the limit of large $m_{\chi_1^\pm}$ the
diphoton signal strength at the intersection point approaches
unity. Moreover, once the chargino spectum is fixed, the maximum
$\mathcal R_{h\gamma\gamma}$ occurs for the lowest possible value of
$m_\Sigma$ because it allows the lowest (highest) possible value of
$\tan\beta_c$ ($\lambda_c$) compatible with the fixed value of $m_h$.

In summary from this brief analysis at low $m_\Sigma$ it seems that
rather large diphoton enhancements can arise in the most natural
frameworks (i.e.~$m_\Sigma$ small, as discussed in
section~\ref{subsec:sector}). However, achieving a firm conclusion on
the naturalness of a large diphoton excess would only be possible by
a numerical study on the tuning needed in the chargino parameters
(constrained by the $T$-parameter bound) and on the low-$m_\Sigma$
phenomenology occurring at $(\lambda,\tan\beta)$ different from the
intersection point.

\section{Conclusion}

In view of the actual value of the Higgs mass provided by the ATLAS
and CMS experiments at the LHC, the MSSM as a solution to the
hierarchy problem loses part of its naturalness (as the theory
requires heavy stops and large mixing) and thus develops a little
hierarchy problem. A way to alleviate this issue is to enlarge
  the MSSM field content by singlets and/or triplets strongly (but
  perturbatively) coupled to the Higgs sector as they introduce
  additional tree-level contributions to the SM-like Higgs mass.
On top of that triplets can
also increase the $\gamma\gamma$ Higgs branching ratio by means of
extra chargino loops in the process $h\to\gamma\gamma$, in agreement
with recent data on diphoton production at the LHC. This fact was
explicitly exhibited in a previous publication~\cite{Delgado:2012sm}
focusing on the decoupling regime, i.e.~assuming the next-to-lightest
CP-even ($H$) and the lightest CP-odd ($A$) Higgses much heavier than
the electroweak scale. In this regime the tree-level couplings in the
decay rates $h\to bb,\,\tau\tau,\, tt,\,WW,\,ZZ$ equal the SM ones and
only the departure of the process $h\to \gamma\gamma$ is
relevant. However, on the one hand the decoupling regime (where $m_A$
is much larger than the electroweak scale) introduces an extra
\textit{inherent fine-tuning} as cancellations between $\mathcal O
(m_A^2)$ terms are required to fix $m_Z^2$ to its experimental
value. On the other hand in the decoupling limit the SM-like Higgs is
indistinguishable from the SM one in the channels dominated by the
tree-level Higgs decays (only channels generated by loop corrections
are modified) while \textit{the extra Higgs sector is very heavy and
  thus difficult to detect}. These features make it interesting to
consider the non-decoupling regime of supersymmetric theories.

In this paper we have considered the non-decoupling regime of the
theory where a zero hypercharge triplet superfield is added to the
MSSM. In this case, in which all the Higgs sector is light and within
the LHC energy range, the issue concerning the phenomenological
feasibility of the model is two-fold:
\begin{enumerate}
\item[\it{i)}] The prediction for $\sigma(pp\to h)BR(h\to XX)$ should be in
  agreement with the experimental data at the LHC. It should then not
  deviate too much from the SM prediction, possibly except for the
  diphoton channel.
\item[\it{ii)}] The prediction for $\sigma(pp\to H)BR(H\to XX)$ should
  explain why the state $H$ (if kinematically accessible at the LHC)
  has not been detected.
\end{enumerate}
Requirement \textit{i)} has been positively fulfilled as in the
considered theory there is, for any value of $m_A$, a SM-like
point at small $\tan\beta$ (whose value depends on the other
supersymmetric parameters) for which all couplings between $h$ and the
SM fields equal the corresponding SM ones. Similar properties also
arise in a second peculiar point which however seems to exist only for
some specific choice of $m_A$ and supersymmetric parameters.  Notably,
the SM-like point does not exist in the MSSM but in principle it does
in theories where singlets and/or triplets with non-zero hypercharge are
introduced.  Therefore all signal strengths of decay rates proceeding
by tree-level diagrams satisfy the condition $\sigma(pp\to h)BR(h\to
XX)\simeq \sigma_{SM}(pp\to h)BR_{SM}(h\to XX)$ and thus they are in
good agreement with experimental data. Large modifications can instead
appear in branching ratios of loop-induced processes, as e.g.~in the
diphoton channel where we can find some large enhancement with respect
to the SM result, depending on the chargino mass spectrum. As the
value of the couplings of $h$ and $H$ to various final states are
correlated we find
strong suppressions of $\sigma(pp\to H)BR(H\to XX)$ which can solve
the issue \textit{ii)} above. In fact we have found that the only
relevant production rates are for $H\to bb,\tau\tau$ which are less
than 10\% the values expected in the SM for a Higgs of the same mass,
and which might eventually be discovered at the high luminosity LHC14 or in a
future linear collider.

Of course once we have opened Pandora's box of a light scalar sector
there are other processes which should be investigated, in particular
those involving the pseudoscalar $A$ and charged $H^\pm$ Higgses,
especially the former which is the lightest one. We have only briefly
discussed some of the possible signatures for production and decay of
$A$ and $H^\pm$, partly due to the absence of experimental data for
these particles. However a dedicated study, both theoretical and
experimental, covering this region is worth in the future when more
data concerning the Higgs sector will be accumulated at the LHC.

\section*{\sc Acknowledgments}
GN and MQ thank the ICTP for hospitality during the first stages
of this work. AD was partly supported by the National Science
Foundation under grant PHY-1215979. MQ was supported by the Spanish
Consolider-Ingenio 2010 Programme CPAN (CSD2007-00042) and by
CICYT-FEDER-FPA2008-01430 and FPA2011-25948.

\section*{Appendix}
\appendix
\section{Analytic expressions for the SM-like point}

At one-loop the matrix $\M^2$ in eq.~\eqref{matrix} can be written as
\be
\mathcal M^2=\left(
\begin{array}{cc}
m_A^2 \cos^2\beta+\wz\sin^2\beta &(-m_A^2+\hz)\sin\beta\cos\beta \\
(-m_A^2+\hz)\sin\beta\cos\beta&m_A^2  
\sin^2\beta+\oz\cos^2\beta 
\end{array}
\right)~,
\ee
where we have used the redefinitions 
\bea
\hz&=&\lambda^2 v^2 - m_Z^2  +\Delta_{\tilde
  t}\M^2_{12}+\Delta_{\Sigma}\M^2_{12}~,\\
\wz&=&m_Z^2+\Delta_{\tilde t}\M^2_{11}+\Delta_{\Sigma}\M^2_{11}~,\\
\oz&=& m_Z^2+\Delta_{\Sigma}\M^2_{22}~.
\eea
It follows that the system of equations \eqref{sistema} turns out to
be 
\bea
\begin{cases}
m_h^4+\cos ^2\beta  \left(m_h^2 (\wz-\oz)+\sin ^2\beta  \left(\wz
   \oz-{m}_{12}^4\right)\right)-m_h^2 \wz=0\\
-m_h^2+\wz \sin ^4\beta +(\oz-2 \hz) \cos ^4\beta +2 \hz \cos ^2\beta =0
\end{cases}
\eea
whence for $\hz>0$ one obtains
\bea
m^2_{12,c}&=&m_h^2+\sqrt{(m_h^2-\wz)(m_h^2-\oz)}~,\nn\\
\cos^2\beta_c&=&\left(1+\sqrt{\frac{m_h^2-\oz}{m_h^2-\wz}}\right)^{-1}~.
\label{solAn}
\eea

Finally one can check analytically that the intersection points in the
left and right panels of Fig.~\ref{fig:1} correspond to the same
parameter point. By using eq.~\eqref{autovalores} to express $\Delta$
in eq.~\eqref{sina}, one observes that at the point \eqref{solAn}
$\sin\alpha$ is independent of $m_A$ and in particular, as it occurs
in the decoupling limit $m_A\to\infty$, it is
$\sin^2\alpha_c=\cos^2\beta_c$.

\end{document}